\documentclass[10pt, conference, letterpaper]{IEEEtran}
\IEEEoverridecommandlockouts
\usepackage{cite}
\usepackage{amsmath,amssymb,amsfonts}
\usepackage{algorithmic,algorithm}
\usepackage{graphicx}
\usepackage{textcomp}
\usepackage{xcolor}
\usepackage{color}
\usepackage{subfigure}
\usepackage{url}
\usepackage{booktabs}
\usepackage{array}
\def\BibTeX{{\rm B\kern-.05em{\sc i\kern-.025em b}\kern-.08em
		T\kern-.1667em\lower.7ex\hbox{E}\kern-.125emX}}
\begin{document}
	
	\title{
		On Designing Multi-UAV aided  Wireless Powered Dynamic Communication via Hierarchical Deep Reinforcement Learning 
	}
	\author{\IEEEauthorblockN{Ze Yu Zhao$^1$, Yue Ling Che$^2$, Sheng Luo$^3$, Gege Luo$^4$, Kaishun Wu$^5$, and Victor C. M. Leung$^6$}
		\IEEEauthorblockA{College of Computer Science and Software Engineering, Shenzhen University, China\\ Email: \{2100271103$^1$, 2110276239$^4$\}@email.szu.edu.cn, \{yuelingche$^2$, sluo$^3$, wu$^5$\}@szu.edu.cn}, vleung@ieee.org$^6$ 
	} 
	\maketitle

	

	\begin{abstract}
		This paper proposes a novel design on the wireless powered communication network (WPCN) in dynamic environments under the assistance of multiple unmanned aerial vehicles (UAVs). Unlike the existing studies, where the low-power wireless nodes (WNs) often conform to the coherent harvest-then-transmit protocol, under our newly proposed double-threshold based WN type updating rule, each WN can dynamically and repeatedly update its WN type as an E-node for non-linear energy harvesting over time slots or an I-node for transmitting data over sub-slots. To maximize the total transmission data size of all the WNs over $T$ slots, each of the UAVs individually determines its trajectory and binary wireless energy transmission (WET) decisions over times slots and its binary wireless data collection (WDC) decisions over sub-slots, under the constraints of each UAV's limited on-board energy and each WN’s node type updating rule. However, due to the UAVs’ tightly-coupled trajectories with their WET and WDC decisions, as well as each WN’s time-varying battery energy, this problem is  difficult to solve optimally. We then propose a new multi-agent based hierarchical deep reinforcement learning (MAHDRL) framework with two tiers to solve the problem efficiently, where the soft actor critic (SAC) policy is designed in tier-1 to determine each UAV’s continuous trajectory and binary WET decision over time slots, and the deep-Q learning (DQN) policy is designed in tier-2 to determine each UAV’s binary WDC decisions over sub-slots under the given UAV trajectory from tier-1. Both of the SAC policy and the DQN policy are executed distributively at each UAV. Finally, extensive simulation results are provided to validate the outweighed performance of the proposed MAHDRL approach over various state-of-the-art benchmarks.

		
	\end{abstract}

	\newtheorem{definition}{\underline{Definition}}[section]
	\newtheorem{fact}{Fact}
	\newtheorem{assumption}{Assumption}
	\newtheorem{theorem}{\underline{Theorem}}[section]
	\newtheorem{lemma}{\underline{Lemma}}[section]
	\newtheorem{corollary}{\underline{Corollary}}[section]
	\newtheorem{proposition}{\underline{Proposition}}[section]
	\newtheorem{example}{\underline{Example}}[section]
	\newtheorem{remark}{\underline{Remark}}[section]
	\newtheorem{observation}{\underline{Observation}}[section]
	\newcommand{\mv}[1]{\mbox{\boldmath{$ #1 $}}}

	\section{Introduction}
	The unmanned aerial vehicle (UAV) aided wireless powered communication network (WPCN) has emerged as one promising technology to realize the energy-sustainable wireless communications. As compared to the traditional WPCNs with fixed deployment on ground \cite{che2015spatial} and \cite{wang2016adaptively}, the UAVs with high mobility are able to provide more efficient radio frequency (RF) wireless energy transmissions (WET) to the low-power wireless nodes (WNs) over the largely shortened energy transmission distances \cite{che2023uav}. This not only evokes an increased number of sufficiently-charged WNs and thus higher communication throughput in UAV-aided WPCNs in general \cite{xu2018uav}, but also enables adaptive and on-demand wireless services in different network scenarios \cite{wang2021chase}. Due to the WNs' dynamically changing demands for energy harvesting and/or information transmissions over time, the design of the UAV-aided WPCN is generally challenging. In the literature, for the ease of the WPCN design, either the WNs conform to the coherent harvest-then-transmit protocol \cite{xie2018throughput}, or the UAVs are pre-assigned to transmit energy or collect wireless data as separate groups \cite{oubbati2022synchronizing}. The dynamics of the UAV-aided WPCN have not been fully explored yet. This motivates us to investigate both of the WNs' time-varying service demands and the UAVs' individually-adaptive trajectory and transmission designs, for enabling the UAV-aided WPCN in dynamic environments.
	
	\subsection{Related work}
	
	\emph{UAV-aided wireless communications}: Benefiting from the air-to-ground (A2G) wireless channels with high quality, the UAVs have been utilized as the aerial base stations (ABS) to provide ubiquitous wireless connections \cite{wang2019multiple}. The line-of-sight (LoS) probability based A2G channel model has been investigated in \cite{haha}, which shows that the LoS probability between the UAV and the WN on ground generally increases with their in-between elevation angle. The joint design of the UAVs' trajectories and communication resource allocations is essential for the UAV-aided wireless communications. For the single-UAV scenario, the UAV's trajectory and time resource allocation are usually alternatively optimized via the successive convex approximation (SCA) technique to, e.g., maximize the WNs' minimum achievable data transmission rate \cite{zhan2017energy} or minimize the age of information (AoI) of the WNs' transmission data \cite{liu2020uav} and \cite{8756665}. The SCA technique has also been employed in the multi-UAV communication scenario \cite{wu2018joint}, but the resultant system performance may be comprised for the high complexity caused by the coupling between the multiple UAVs' trajectories and transmissions \cite{hassan20223to}. To achieve higher downlink capacity, the deep reinforcement learning (DRL) is leveraged in \cite{9931560} and \cite{wang2019multi} to determine the wireless data collection (WDC) schemes and the movements of multiple UAVs.

	\emph{UAV-aided WET}: The feasibility of the UAV-aided WET has been validated by the field tests in \cite{clerckx2018fundamentals}. The joint design of the UAV trajectory and WET has been widely studied in \cite{shi2023two,ren2022energy,baek2019optimal} for the single-UAV scenario or in \cite{mu2022trajectory} and \cite{9468714} for the multi-UAV scenario, respectively, to maximize the total harvested energy at the WNs. However, all the above studies ideally assumed that all the WNs not only have the same energy demands, but also utilize linear energy harvesters, where each WN harvests non-zero direct current (DC) energy regardless of the UAVs' transmission distances. It is noted that for multi-UAV aided WET, \cite{magrl} proposed a new metric called the hungry-level of energy (HoE) to measure the WNs' different and time-varying energy demands, and exploited the DRL algorithm to design on-demand WET based on non-linear energy harvesting at the WNs.

	\emph{UAV-aided WPCN}: Despite of the extensive studies on the UAV-aided communications and the UAV-aided WET, the design of the UAV-aided WPCN is not a simple combination of them. Since the WNs use their harvested energy from the UAVs' WET in the downlink to support their data transmissions in the uplink to the UAVs, the UAVs' WET performance may bottleneck the WNs' data transmissions in the UAV-aided WPCN. In the literature, the UAVs' task period is usually divided into a WET phase and a follow-up WDC phase \cite{liu2021average,luo2020joint,che2020uav}, where each phase has the same time length for all the UAVs. While such a two-phase protocol simplifies the UAV-aided WPCN design, due to the WNs' generally different energy demands, not all the WNs can harvest sufficient energy in the UAVs' WET phase, and thus not all the WNs' data transmission requirements can be satisfied in the UAVs' WDC phase. Moreover, in \cite{oubbati2022synchronizing}, instead of evoking adaptive WET and WDC at each of the UAVs, multiple UAVs were divided into two groups to separately transmit energy to or collect data from the WNs, where the deep deterministic policy gradient (DDPG) algorithm is applied to jointly design the multi-UAV's trajectory and resource allocations.

	\subsection{Our Contributions}
	
	In this paper, we propose a novel design of the wireless powered dynamic communication network under the assistance of multiple UAVs, where each of the WNs with different energy demands can dynamically harvest energy whenever in short energy supply and transmit data whenever sufficiently charged. To adapt to the WNs' different service demands, each of the UAVs individually determines its trajectory and binary WET decisions over time slots and its binary WDC decisions over sub-slots. However, due to the UAVs' tightly-coupled trajectories with their WET and WDC decisions, as well as each WN's time-varying battery energy, the optimal network design is very challenging to realize using the traditional optimization techniques.
	By taking each UAV as an individual agent, we then propose a novel multi-agent based hierarchical DRL (MAHDRL) framework with two tiers, to address the UAVs' action decisions distributively over the two different time scales of time slots and sub-slots.
	
	Our main contributions are summarized as follows: 
	
	\emph{1) Practical System Model with Repeatedly Updating WN Types}: Section \ref{section: system model} models the multi-UAV aided WPCN by exploiting the LoS-probability based A2G channel model. Under the newly proposed double-threshold based WN type updating rule, each WN repeatedly updates its node type as an E-node for harvesting energy or an I-node for transmitting data over time slots. In each time slot, if a WN is an E-node, a practical non-linear energy harvesting model is applied for its energy harvesting; and if a WN is an I-node, the sub-slot based data transmission scheduling is used to alleviate the UAVs' received co-channel interference. The battery energy management at each WN and each UAV are all addressed.
	
	\emph{2) Novel Problem Formulaion Solved by MAHDRL Framwork}: Section \ref{section: problem formulation} formulates the problem to maximize the total transmission data size of all the WNs over $T$ slots, by jointly optimizing the trajectory, the WET decisions, and the WDC decisions of each UAV, under the constraints of each UAV's limited on-board energy and each WN's node type updating rule. Due to the high complexity to solve this problem optimally, we propose the novel MAHDRL framework with two tiers to solve the problem efficiently, where the soft actor critic (SAC) policy is executed in tier-1 to determine each UAV's continuous trajectory and binary WET decision over time slots, and the deep-Q learning (DQN) policy is executed in tier-2 to determine each UAV's binary WDC decisions over sub-slots under the given UAV trajectory from tier-1.
	
	\emph{3) Hierarchical Training of the Central SAC and the Local DQN}: Sections \ref{section: SAC} designs the central training of the SAC policy for tier-1, due to its relatively higher complexity caused by the entropy-based action space exploration, and Section \ref{section: DQN} designs the local training of the DQN policy with lower complexity in tier-2, respectively. For both policies, due to each E-node's low battery energy and thus weak transmit power to report its status, only UAVs nearby (if any) can observe with its actual status, which generally leads to the partially observable Markov decision process (POMDP) based modeling at both the central trainer and the individual UAV. Due to the mutually affected policy decisions among the two tiers, the reward functions for both the SAC and the DQN are designed to achieve the proper trade-off between satisfying the E-nodes' and the I-nodes' different service demands under the WN type updating rule. Once well trained, both of the SAC policy and the DQN policy are executed distributively at each UAV.
	
	\emph{4) Extensive simulation results for Performance Evaluation}: Section \ref{section: simulation} validates the outweighed performance of the proposed MAHDRL approach over various state-of-the-art benchmarks in both of the training stage and the test stage. A network example is also illustrated to elaborate the dynamics of the WNs' node type variations and the UAVs' adaptive trajectories. Moreover, the scalability of the proposed MAHDRL approach in different network scales is also validated.

	\begin{figure}
		\setlength{\abovecaptionskip}{-0.0in}
		\centering
		\DeclareGraphicsExtensions{.eps,.mps,.pdf,.jpg,.png}
		\DeclareGraphicsRule{*}{eps}{*}{}
		\includegraphics[angle=0, width=0.49 \textwidth]{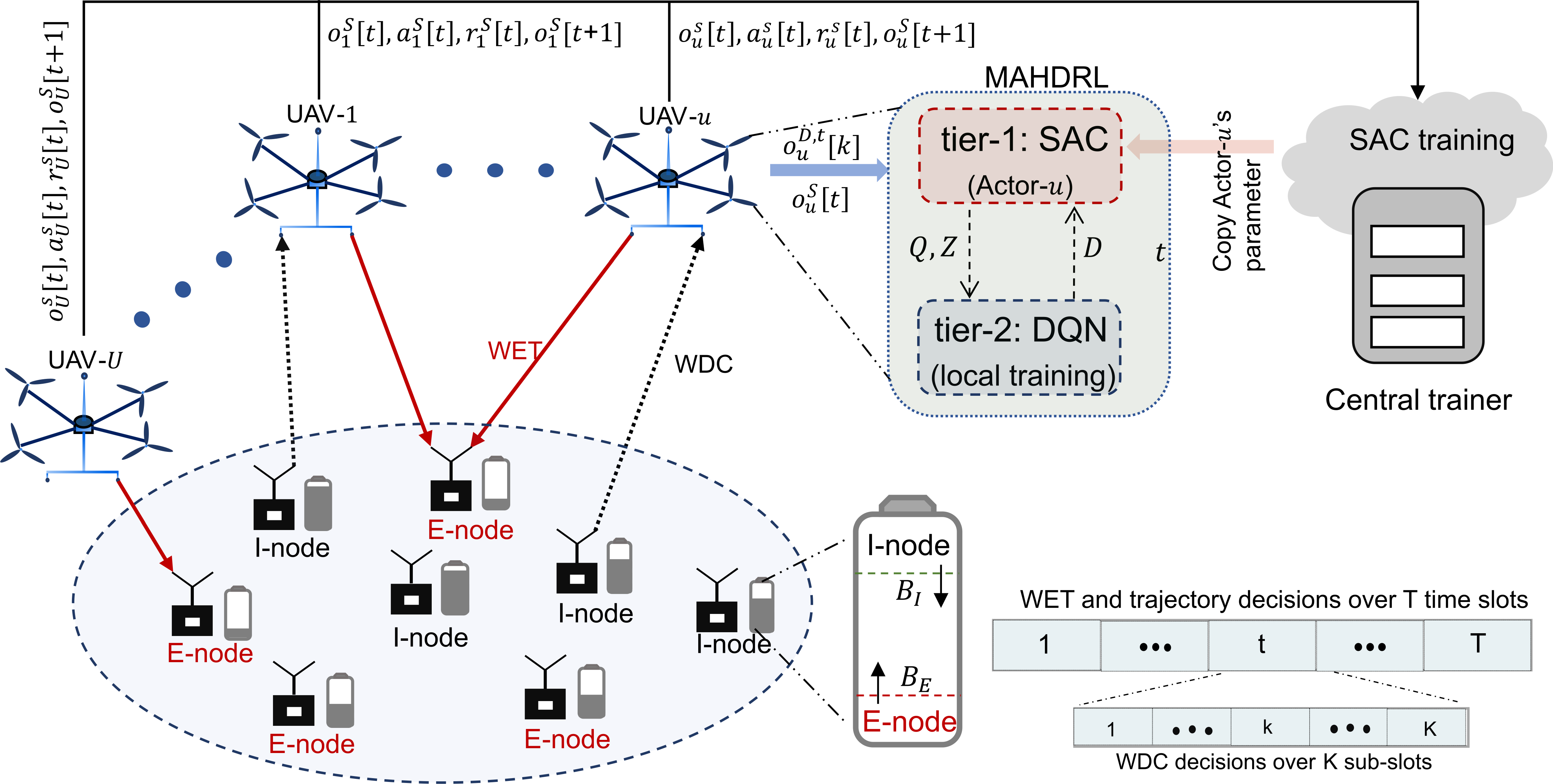}
		\caption{Multi-UAV aided WPCN with repeatedly changing WN types.}
		\label{fig: system model}
		\vspace{-0.25in}
	\end{figure}
	
	\section{System Model}\label{section: system model}

	As shown in Fig.~\ref{fig: system model}, we consider a multi-UAV aided WPCN, where each of the $U\ge 2$ UAVs transmits wireless energy to and/or collects wireless data from in total of $W$ WNs for a task period of $T$ slots with $\mathcal{T}=\{1,...,T\}$, $\mathcal{U}=\{1,...,U\}$ and $\mathcal{W}=\{1,...,W\}$. 
	Each UAV flies at a fixed altitude of $h$ meters (m). To prevent the UAVs' WET in the downlinks from causing co-channel interference to their WDC in the uplinks, each UAV is equipped with two antennas to enable separate WET and WDC over different and non-overlapped frequency bands at the same time.
	
	Each WN is installed with a single antenna and a rechargeable battery, and either harvests energy from or transmits data to the UAVs in each slot.
	The WNs that harvest energy in slot $t$ are referred to as the E-nodes. Each of the E-nodes stores the harvested energy in its rechargeable battery. The WNs that transmit data in slot $t$ are referred to as the I-nodes. All the I-nodes in slot $t$ have sufficient battery energy to support their data transmissions. We use $F_w [t]=\{0,1\}$ to label the type of WN-$w$ in slot $t\in \mathcal{T}$, where WN-$w$ is an I-node if $F_w[t]=1$, or an E-node, otherwise. Correspondingly, the WN set $\mathcal{W}$ is divided into the I-node subset $\mathcal{I}[t] \triangleq\{w|F_w[t]=1, w\in \mathcal{W}\}$ and the E-node subset $\mathcal{E}[t] \triangleq\{w|F_w[t]=0, w\in \mathcal{W}\}$ in slot $t$ with $\mathcal{W} = \mathcal{E}[t] + \mathcal{I}[t]$. Due to the time-varying WN battery energy, each WN's type and thus the elements in $\mathcal{E}[t]$ and $\mathcal{I}[t]$ may all change over different time slots. TABLE \ref{table: notations} gives the notations of the key parameters in this paper.
	\begin{table}[h]
		\centering
		\caption{ Notations of key parameters }\label{table: notations}
		\scriptsize
		\setlength{\tabcolsep}{1.3 mm}{
			\begin{tabular}{c  c}
				\toprule[1.5pt]
				\textbf{Notations}     & \textbf{Description} \\ \midrule[1pt]
				$F_w[t]$ & The type of WN-$w$ in slot $t$   \\
				\specialrule{0em}{1pt}{1pt}
				$q_u[t]$, $q_w$  & Location of UAV-$u$ or WN-$w$ in slot $t$ \\  
				\specialrule{0em}{1pt}{1pt}
				$d_w^u[t]$  &    Distance between UAV-$u$ and WN-$w$ in slot $t$   \\ 
				\specialrule{0em}{1pt}{1pt}
				$\vartheta$, $\varrho$      &  Time length of a slot or a sub-slot \\
				\specialrule{0em}{1pt}{1pt}
				$P_W$, $P_U$  &   Each WN's or UAV's transmit power \\
				\specialrule{0em}{1pt}{1pt} 
				$G_w^u[t]$ &    Channel gain between WN-$w$ and UAV-$u$ in slot $t$    \\
				\specialrule{0em}{1pt}{1pt}
				$Z_u[t]$ & UAV-$u$’s WET decision in slot $t$ \\
				\specialrule{0em}{1pt}{1pt}
				$E_w^{har}[t]$ &  Harvested energy at WN-$w$ in slot $t$   \\
				\specialrule{0em}{1pt}{1pt}
				$D_{u,w}^t[k]$  & UAV-$u$'s WDC decison for WN-$w$ at sub-slot $k$ in slot $t$   \\
				\specialrule{0em}{1pt}{1pt}
				$M_{u,w}^t[k]$  & WN-$w$’s transmission data size to UAV-$u$ at sub-slot $k$ in slot $t$ \\  
				\specialrule{0em}{1pt}{1pt}
				$C_w[t]$  &  Aggregated transmission data size of WN-$w$ in slot $t$   \\
				\specialrule{0em}{1pt}{1pt}
				$B_w[t]$, $B_u[t]$ &  Battery level of WN-$w$ or UAV-$u$ in slot $t$    \\
				\bottomrule[1.5pt]
		\end{tabular}}
	\end{table}

	\subsection{LoS-Probability based Channel Model}
	
	Denote the coordinate of WN-$w$ as $q_w=(x_w,y_w,0)$ and that of UAV-$u$ in slot $t$ as $q_u [t]=(x_u [t],y_u [t],h)$, respectively, $\forall w\in \mathcal{W}$ and $\forall u\in \mathcal{U}$. Since the time length $\vartheta$ of each slot is generally very short, each UAV's location is assumed to be unchanged within each slot. The distance between UAV-$u$ and WN-$w$ in  slot $t$ is obtained as $d_w^u[t]= \left \| q_w-q_u[t] \right\|$, where $\left\|\cdot\right\|$ is the Euclidean norm.
	
	Let $P_{LoS,w}^u[t] \!=\! \left( 1+a\exp \left( -b(\beta_w^u[t]-a \right)) \right)^{-1}$ denote the LoS probability between UAV-$u$ and WN-$w$ in slot $t$ \cite{haha}, where $\beta_w^u[t]=\sin^{-1}\left( h/d_w^u[t]\right)$ is their in-between elevation angle in slot $t$, and $a$ and $b$ are two constant parameters measured from the environment. The non-line-of-sight (NLoS) probability is thus $P_{NLoS,w}^u [t] = 1 - P_{LoS,w}^u [t]$. The average channel gain between UAV-$u$ and WN-$w$ in slot $t$ is obtained as
	\small
	\begin{equation} 
		\label{equ: channel gain}
		G_w^u[t] = P_{LoS,w}^u[t]G_0d_w^u[t]^{-\alpha_L} + P_{NLoS,w}^u[t]G_0d_w^u[t]^{-\alpha_N},
	\end{equation}
	\normalsize
	where $G_0$ is the average channel gain at a reference distance of $1$ m, and $\alpha_L$ and $\alpha_N$ with $0<\alpha_L<\alpha_N$ are the LoS link and NLoS links' path-loss exponents, respectively.
	
	\subsection{UAVs' Energy Transmissions to E-nodes}
	Denote $Z_u[t]\in \{0,1\}$ as UAV-$u$’s WET decision in slot $t$, where UAV-$u$ transmits energy with a fixed transmit power $P_U>0$ in slot $t$ if $Z_u[t]=1$, or keeps silent on the frequency band for WET, otherwise, to save its limited on-board energy. We consider a non-linear energy harvester at each E-node, which transforms its received RF power $p$ at the antenna into the DC power $\bar{P}(p)$ stored in the battery nonlinearly as follows \cite{alevizos2018sensitive}:
	\small
	\begin{equation}
		\label{equ: Energy Harvesting Conversion}
		\bar{P}(p)=\begin{cases}
			0,& p\in [0, P_{sen}), \\
			f(p),& p\in[P_{sen}, P_{sat}), \\
			f(P_{sat}),& p \in [P_{sat}, +\infty],
		\end{cases}
	\end{equation}
	\normalsize
	where $P_{sen}$ and $P_{sat}$ with $0<P_{sen}<P_{sat}$ are the sensitivity power and the saturation power of the energy harvester, respectively, and $f(\cdot)$ is a non-linear power transform function that can be obtained through the curve fitting technique \cite{alevizos2018sensitive}. From (\ref{equ: Energy Harvesting Conversion}), no DC power is harvested if $p<P_{sen}$ and the harvested DC power keeps unchanged if $p \ge P_{sat}$. Since the received RF power at E-node-$w$ from all the UAVs is $\sum_{u=1}^{U}P_UZ_u[t]G_w^u[t]$, its harvested energy in slot $t$ is
	\vspace{-0.07in}
	\small
	\begin{equation}
		\label{equ: harvested energy}
		E_w^{har} [t]=\bar{P} \left (\sum_{u=1}^{U}P_UZ_u[t]G_w^u[t] \right) \vartheta, \forall w \in \mathcal{W}.
	\end{equation}
	\normalsize
	Due to the practically high value of $P_{sen}$ (with, e.g., -10 dBm \cite{cpdd}), the UAVs with $Z_u[t]=1$ need to locate close to E-node-$w$ to assure its non-zero energy harvesting. 
	
	Denote $B_w [t]$ as the battery energy level of WN-$w\in \mathcal{W}$ at the begining of slot $t\in\mathcal{T}$ and $B_W^{max}$ as the WN battery capacity, respectively. If WN-$w$ is an E-node in slot $t$, i.e., $F_w[t]=0$, based on (\ref{equ: harvested energy}), $B_w[t+1]$ is updated as 
	\small
	\begin{equation}
		\label{equ: E-node battery update}
		B_w[t \!+\!1]=
		\min \left(B_W^{max}, B_w[t]+E_w^{har}[t] \right), \forall w \in \mathcal{E}[t].
	\end{equation}
	\normalsize
	
	\subsection{UAVs' Data Collections from I-nodes}
	
	For the I-nodes, to alleviate the UAVs' received co-channel interference in each time slot, we increase the selections of time resources by dividing each time slot into $K\ge 2$ sub-slots, as shown in Fig.~\ref{fig: system model}. Each sub-slot is of time length $\varrho$ with $\vartheta=K\varrho$. Denote $D_{u,w}^t [k] \!\in\! \{0,1\}$ as UAV-$u$'s sub-slot based WDC decision, where $D_{u,w}^t [k]=1$ represents that UAV-$u$ collects data from I-node-$w$ at the $k$-th sub-slot in slot $t$, or $D_{u,w}^t [k] \!=\!0$, otherwise. It is considered at any sub-slot $k$, each UAV collects data from at most one I-node, and each I-node transmits data to at most one UAV, which is expressed as
	\small
	\begin{equation}
		\label{constraint: scheduling}
		\sum_{w=1}^{W}D_{u,w}^t[k] \leq 1  
		\textrm{~and~}
		\sum_{u=1}^{U}D_{u,w}^t[k] \leq 1, ~\forall u\in \mathcal{U},\forall w\in \mathcal{I}[t].
	\end{equation}
	\normalsize
	
	Denote $\Gamma_{u,w}^t[k]$ as the signal-to-interference-plus-noise-ratio (SINR) received at UAV-$u$ from I-node-$w$ at the $k$-th sub-slot in slot $t$. Denote $P_W>0$ as each I-node's fixed transmit power. Due to the sufficiently-short time slot length, $G_w^u[t]$ in (\ref{equ: channel gain}) is assumed to remain unchanged over all sub-slots in slot $t$. As a result, $\Gamma_{u,w}^t[k]$ is obtained as
	\small
	\begin{equation}
		\label{equ: sinr}
		\Gamma_{u,w}^t[k] \! =\! \frac{D_{u,w}^t[k]P_WG_w^u[t]}{\sum_{i\in\mathcal{U}} \! \sum_{j \! \in \! \mathcal{I}[t],j \! \neq \! w} \! D_{i,j}^t[k] \! P_W \! G^{u}_{j}[t] \! + \! \sigma^2},
	\end{equation}
	\normalsize
	where $\sigma^2$ is the received noise power at each UAV. Denote $M_{u,w}^t[k]$ as I-node-$w$'s transmission data size (bits/Hz) to UAV-$u$ at the $k$-th sub-slot in slot $t\in \mathcal{T}$. We have $M_{u,w}^t[k] = \log_2\left( 1+\Gamma_{u,w}^t[k] \right)\varrho$. Hence, the aggregated transmission data size of I-node-$w$ over all the $K$ sub-slots in slot $t$, denoted by $C_w[t]$, is obtained as
	\small
	\begin{equation} 
		\label{equ: unit instantaneous rate}
		C_w[t] = \sum_{u=1}^{U}\sum_{k=1}^{K}M_{u,w}^t[k].
	\end{equation}
	\normalsize
	Denote $C_{total}$ as the total transmission data size of all the I-nodes over all the $T$ slots with $C_{total}=\sum_{t=1}^{T}\sum_{w=1}^{W}C_w[t]$.

	To support the data transmissions, each I-node-$w$ consumes an energy amount of $\sum_{u=1}^{U}\sum_{k=1}^{K}D_{u,w}^{t}[k]P_W\vartheta$ in slot $t$ from its battery. Thus, if WN-$w$ is an I-node with $F_w[t]=1$, for a given $B_w[t]$, $B_w[t+1]$ is updated as 
	\small
	\begin{equation}
		\vspace{-0.00in}
		\label{equ: I-node battery update}
		B_w[t+1]\!=\!
		\max \! \left(\! B_w[t] \!- \! \sum_{u=1}^{U}\!\sum_{k=1}^{K}D_{u,w}^{t}[k]P_W\vartheta, \! 0\! \right),\! \forall w \! \in \! \mathcal{I}[t].
	\end{equation}
	\normalsize
	
	\subsection{Double-Threshold based WN Type Updating}
	
	At the beginning of each slot, based on each WN is an E-node or I-node in the previous slot, each WN updates its battery energy level according to (\ref{equ: E-node battery update}) or (\ref{equ: I-node battery update}), respectively. Based on the $B_w[t]$, WN-$w$ updates its WN type at the beginning of slot $t$ according to a double-threshold based WN type updating rule given as follows:
	\small
	\begin{equation}
		\label{equ: wireless node flag update}
		F_w[t]\!= \! \begin{cases}
			1, &if~ B_w[t] \! \ge \!  B_I ~~or~ B_E \! < \! B_w[t] \! < \! B_I,~F_w[t \! - \! 1] \! = \! 1,\\ 
			0, &if~ B_w[t] \! \leq \!  B_E ~~or~ B_E \! < \! B_w[t] \! < \! B_I,~F_w[t \! - \! 1] \! = \! 0,
		\end{cases}
	\end{equation}
	\normalsize
	where $B_E$ and $B_I$ are the given E-node threshold and the I-node threshold, respectively, with $P_W\vartheta < B_E < B_I$. From (\ref{equ: wireless node flag update}), WN-$w$ becomes an E-node in slot $t$ when $B_w[t]\leq B_E$, and an I-node in slot $t$ when $B_w[t]\geq B_I$. When $B_E< B_w[t]< B_I$, there are two cases: 1) if WN-$w$ is an I-node in the previous slot $t-1$ with $F_w[t \!-\! 1]=1$, since $B_w[t]$ is still higher than the E-node threshold $B_E$, it keeps transmitting data as an I-node in slot $t$; and 2) if WN-$w$ is an E-node in the previous slot $t \!-\!1$ with $F_w[t \!-\! 1]=0$, since $B_w[t]$ does not exceed the I-node threshold $B_I$, WN-$w$ continues harvesting energy as an E-node in slot $t$.  
	
	It is noted that unlike the widely-used single threshold to determine each WN's type in, e.g., \cite{che2015spatial}, where each WN's data transmission is often suspended due to the frequently-changed node type, by consuming the battery energy from higher than $B_I$ to lower than $B_E$ with a sufficiently large $B_I-B_E$ under (\ref{equ: wireless node flag update}), each I-node's data transmission becomes more reliable.

	\subsection{UAVs’ Energy Consumption Model} \label{subsection: UAVs' Energy Consumption Model}
	The energy consumption of each UAV is mainly caused by the UAV's movement, WET and WDC. Denote $P_I$ as the fixed WDC power consumption at each UAV. The total WDC energy consumption of UAV-$u$ in slot $t$ is $\sum_{w=1}^{W}\sum_{k=1}^{K}D_{u,w}^t[k]P_I\varrho$. From \cite{zeng2019energy}, UAV-$u$'s propulsion power consumption in slot $t$ is determined by its velocity $V_u[t] \triangleq \frac{1}{\vartheta}\left\| q_u[t+1]-q_u[t] \right\|$ as follows:
	\small
	\vspace{-0.1in}
	\begin{align}
		\label{equ: UAV Propulsion Power}
		P_{pro}(V_u[t])&=P_a\left(1+\frac{3V_u[t]^2}{V_{tip}^2}\right )+ \frac{1}{2}f_0\varpi e_1AV_u[t]^3  \nonumber  \\ 
		&+P_b\left (\sqrt{1+\frac{V_u[t]^4}{4e_0^4}}-\frac{V_u[t]^2}{2e_0^2}\right )^{\frac{1}{2} },
		\vspace{-0.1in}
	\end{align}
	\normalsize
	where the details of the constant parameters $P_a $, $P_b$, $ V_{tip}$, $e_0$, $f_0$, $\varpi$, $e_1$ and $A$ are given in \cite{zeng2019energy}. The propulsion energy consumption of UAV-$u$ in slot $t$ is obtained as $P_{pro}(V_u [t])\vartheta$. The WET energy consumption of UAV-$u$ in slot $t$ is obtained as $Z_u [t] P_U \vartheta$. Hence, UAV-$u$'s total energy consumption in slot $t$ is $E_u[t]=\sum_{w=1}^{W}\sum_{k=1}^{K}D_{u,w}^t[k]P_I\varrho+P_{pro}(V_u [t])\vartheta+Z_u [t] P_U \vartheta$. Denote $B_u [t]$ as the battery energy level of UAV-$u$ at the beginning of slot $t\in\mathcal{T}$. We obtain that
	\vspace{-0.1in}
	\small
	\begin{equation}
		\label{equ: UAV battery update}
		B_u[t]  =  \max  \left(B_u[t - 1] -  E_u[t -1] ,  0 \right).
		\vspace{-0.1in}
	\end{equation}
	\normalsize
	Assume that each UAV is fully charged at the initial with $B_u[0]=B_U^{max}$, where $B_U^{max}$ is the UAV battery capacity. At the end of the last slot $t=T$, UAV-$u$'s remained battery energy, denoted by $B_u^{end}$, is obtained by substituting $t \!=\! T + 1$ into (\ref{equ: UAV battery update}).

	\vspace{-0.15in}
	\section{Problem Formulation and MAHDRL framework} \label{section: problem formulation}
	\vspace{-0.1in}
	\subsection{Problem Formulation}
	We jointly optimize the trajectories $\boldsymbol{Q}=\{q_u[t]\}$, the WET decisions $\boldsymbol{Z}=\{Z_u[t]\}$, and the WDC decisions $\boldsymbol{D}=\{D_{u,w}^t[k] \}$ of all the UAVs, to maximize the WNs' total transmission data size $C_{total}$, subject to the constraints on each WN's minimum required transmission data size, and each WN's battery energy and WN type variations over time, as well as each UAV's trajectory and battery energy constraints. The problem is formulated as (P1).
	
	In problem (P1), the constraint in (\ref{constraint: a}) guarantees that the overall transmission data size of each WN in $T$ slots is no smaller than the minimum required data size  $C_{min}>0$; the constraint in (\ref{constraint: b}) ensures that each UAV's velocity does not exceed the maximum allowable velocity $V_U^{max}$; the constraint in (\ref{constraint: c}) guarantees that UAV-$u$'s remained battery energy at the end is not lower than a required level $B_U^{min}$ for, e.g., its safe return; the constraints in (\ref{constraint: D}) and (\ref{constraint: d}) represent the binary WDC and WET decisions of each UAV, respectively; the constraint in (\ref{constraint: e}) assures a safe distance of $d_{min}$ between any two UAVs.
	\small
	\begin{align} 
		\textrm{(P1)}:~\max_{\boldsymbol{Q},\boldsymbol{Z},\boldsymbol{D}}&~~  
		\sum_{t=1}^{T}\sum_{w=1}^{W}C_w[t] \nonumber \\
		\mathrm{s.t.}  ~~&(\ref{equ: E-node battery update}),~(\ref{constraint: scheduling}),~(\ref{equ: I-node battery update}),~(\ref{equ: wireless node flag update}),~(\ref{equ: UAV battery update}), \nonumber \\
		& \sum_{t=1}^{T}C_w[t] \ge C_{min}, \forall w \in \mathcal{W},  \label{constraint: a} \\
		& V_u[t] \! \leq \!  V_U^{max}, \forall u \in \mathcal{U}, \forall t \in \mathcal{T},  \label{constraint: b} \\
		&B_u^{end}\ge B_U^{min}, \forall u \in \mathcal{U}, \label{constraint: c} \\
		& D_{u,w}^t[k] \! \in \! \{0,1\}, \forall u\in\mathcal{U}, \forall w \in \mathcal{I}[t], \forall t \in \mathcal{T},  \label{constraint: D} \\
		& Z_u[t]\in\{0,1\}, \forall u\in \mathcal{U}, \forall t \in \mathcal{T}, \label{constraint: d} \\
		&d_u^{u^{'}}[t] \! \ge \!  d_{min}, ~\forall u, u^{'} \in \mathcal{U}, u \!  \neq \!  u^{'}, \forall t \in \mathcal{T},  \label{constraint: e}
		\vspace{-0.0in}
	\end{align}
	\normalsize
	
	Problem (P1) is a very complicated mixed-integer programming problem, which is difficult to solve optimally. In particular, to maximize $C_{total}$, the WNs are expected to become the I-nodes to transmit data in most of the $T$ slots based on (\ref{equ: wireless node flag update}), which in turn requires each E-node to harvest sufficient energy rapidly. However, due to the E-nodes' different battery energy levels and thus different amounts of energy required to harvest to reach the threshold $B_E$ in (\ref{equ: wireless node flag update}), it is difficult to ensure that most of the E-nodes along each UAV's trajectory can become the I-nodes in each slot. 
	Moreover, while the multiple UAVs with binary WET decisions are encouraged to stay close to enhance an E-node's harvested energy, due to the additive property of the harvested energy in (\ref{equ: harvested energy}), they may also be required to stay far away from each other to alleviate their received co-channel interference for WDC. As a result, the multi-UAVs'  decisions on their trajectories and the binary WET and WDC selections are all tightly coupled over time under the time-varying WN types. In addition, all the UAVs must use their limited on-board energy carefully to meet (\ref{constraint: b}) and (\ref{constraint: e}). Therefore, problem (P1) is very challenging to solve.
	
	\vspace{0.1in}
	\subsection{Proposed MAHDRL Framework} \label{subsection: mahdrl solution}
	\vspace{0.1in}
	Given the high complexity to solve problem (P1), the DRL approach is leveraged. Since there lacks a central controller to determine all the UAVs' $\boldsymbol{Q}$, $\boldsymbol{Z}$ and $\boldsymbol{D}$, we consider a multi-agent based DRL approach with each UAV as an agent \cite{dai2022aoi,wang2021deep,peng2020multi}. However, it is noted that the existing multi-agent based DRL approach using a single DRL algorithm (e.g., DQN \cite{dqn}) may not be able to solve the proposed problem (P1) properly, mainly due to the following three reasons: 
	1) Problem (P1) involves UAV-$u$'s policy decisions over two different time scales, where one is over the time slots for determining $\{Z_u[t]\}$ and $\{q_u[t]\}$, and the other is over the sub-slots for determining $\{D_{u,w}^t[k]\}$, and thus requires at least two DRL algorithms, each for one time scale;
	2) Since a more rapid decision is required for $\{D_{u,w}^t[k]\}$ over the sub-slots than $\{Z_u[t]\}$ and $\{q_u[t]\}$ over the time slots, a lower DRL algorithm complexity is required for the time scale of sub-slots than that over the time slots;
	3) Although $Z_u[t]$ and $\{D_{u,w}^t[k]\}$ are binary, the trajectory  $q_u[t]$ is continuous. This requires the DRL algorithm working over the time slots to have continuous action outputs. 
	
	\vspace{-0.16in}
	To meet the above requirements, a multi-agent based two-tier hierarchical DRL design is proposed to solve problem (P1). As shown in Fig.~\ref{fig: system model}, in the two-tier model at each UAV-$u$, the SAC algorithm \cite{sac} with continuous action outputs is employed in tier-1 to determine UAV-$u$'s trajectories and WET decisions over slots; and in tier-2, based on the determined trajectory in tier-1, the DQN with discrete action outputs is applied to determine UAV-$u$'s WDC decisions over sub-slots. Moreover, although the off-line trained SAC policy is distributively executed at each UAV at tier-1 in the implementation stage, due to the higher complexity of the SAC algorithm than the DQN, as well as the UAVs' mutually coupled decisions on $\boldsymbol{Q}$, $\boldsymbol{Z}$ and $\boldsymbol{D}$ in problem (P1), the SAC is trained centrally (at, e.g., the central trainer) by using the observations gathered from all the UAVs in the training stage. The DQN in tier-2 is both trained and executed at each UAV distributively. As will be specified in the following two sections, since the UAVs may not be able to obtain all the WNs' status, both the SAC and the DQN adopt the POMDP-based modeling.
	
	\section{SAC in Tier-1} \label{section: SAC}
	
	\subsection{Observation and POMDP State} \label{subsection: sac pomdp}
	As shown in Fig. \ref{fig: system model}, the central trainer gathers the local observations of all the UAVs for the SAC's training.
	Each UAV observes the WNs' status from their periodic status reporting at the beginning of each slot $t$ via a common ground-to-air channel. 
	For each E-node-$w$, it reports its current battery energy level $B_w[t]$ and the accumulated transmission data size $\sum_{t^{'}=1}^{t-1}C_w[t^{'}]$ from the first slot to the previous slot $t-1$ to all the UAVs. However, due to its very limited battery energy and thus weak transmit power for status reporting, only (if any) UAVs that locate within a given horizontal distance of $d_{\textrm{cov}} \!>\! 0$ to E-node-$w$ can receive its reporting. Hence, depending on their horizontal distance $d_{rep, w}^u[t]=\sqrt{(x_u[t]-x_w)^2+(y_u[t]-y_w)^2}$ in slot $t$, the observation of E-node-$w$'s battery level at UAV-$u$, denoted by $B_w^u[t]$, is given as
	\small
	\begin{equation}
		\vspace{-0.00in}
		B_w^u[t]= \begin{cases} \label{equ: uav knows battery}
			B_w[t], & if~ d_{rep, w}^u[t] \leq d_{\textrm{cov}}, \\
			B_w^u[t \!- \! 1], &otherwise,
		\end{cases}, \forall w\in \mathcal{E}[t],
	\end{equation}
	\normalsize
	where if $d_{rep, w}^u[t]\leq d_{\textrm{cov}}$, $B_w^u[t]$ is updated as E-node-$w$'s actual battery energy $B_w[t]$ or keeps unchanged as $B_w^u[t \!-\! 1]$. Due to the generally high power sensitivity $P_{sen}$ in (\ref{equ: Energy Harvesting Conversion}), if $d_{rep, w}^u[t] > d_{\textrm{cov}}$, it is assumed that E-node-$w$ harvests zero energy from UAV-$u$ in slot $t$, i.e., $\bar{P}(P_UZ_u[t]G_w^u[t])\vartheta=0$. Similarly, denote UAV-$u$'s observation of E-node-$w$'s accumulated transmission data size by $C_w^u[t]$, which is obtained as
	\small
	\begin{equation}
		\vspace{-0.00in}
		C_w^u[t]= \begin{cases} \label{equ: uav knows data}
			\sum_{t^{'}=1}^{t-1}C_w[t^{'}], & if~ d_{rep, w}^u[t] \leq d_{\textrm{cov}}, \\
			C_w^u[t \!- \! 1], &otherwise,
		\end{cases}, \forall w\in \mathcal{E}[t].
	\end{equation}
	\normalsize
	For each I-node-$w$, besides $B_w[t]$ and $\sum_{t^{'}=1}^{t-1}C_w[t^{'}]$, it also reports its current WN type $F_w[t]=1$ to all the UAVs. Due, to their sufficient battery energy to support the status reporting, it is assumed that each I-node's reported information can be well received at all the UAVs. Hence, for any UAV-$u$, we have $B_w^u[t]=B_w[t]$ and $C_w^u[t]=\sum_{t^{'}=1}^{t-1}C_w[t^{'}]$, $\forall w\in \mathcal{I}[t]$. Considering the very short reporting time period, we assume that only trivial amount of energy is consumed at each E-node and I-node for status reporting, and thus is ignored in (\ref{equ: E-node battery update}) and (\ref{equ: I-node battery update}), respectively, for their battery energy updating\footnote{For the I-nodes, we consider that $B_I$ in (\ref{equ: wireless node flag update}) is generally much higher than the accumulated status reporting energy consumption over all $T$ slots; For the E-nodes, if E-node-$w$'s battery level becomes not sufficient for reporting status in slot $t$, it keeps silent, and all the UAVs update $B_w^u[t]=B_w^u[t-1]$.}.
	
	At each UAV agent, based on its received reporting from the WNs at the beginning of each slot $t$, the observation of UAV-$u$ in slot $t$ is given as $o_u^S [t]=\{F_1 [t],...,F_W [t],B_1^u [t],...,B_W^u [t],$ $C_1^u [t],...,C_W^u [t],x_u [t],y_u [t], B_u [t]\}$, which consists of all the WNs' actual node types in slot $t$, all the WNs' battery energy levels observed at UAV-$u$, all the WNs' accumulated transmission data size observed at UAV-$u$, and UAV-$u$'s own horizontal location and battery energy in slot $t$. Denote $\mathcal{O}^S$ as the observation space with $o_u^S[t] \in \mathcal{O}^S$. Specifically, for the observation of $\{F_w[t]\}$, UAV-$u$ can easily identify the I-node set $\mathcal{I}[t]$ as the set of WNs who report $F_w[t]=1$ in slot $t$ and thus obtains the E-node set as $\mathcal{E}[t] \! =\! \mathcal{W}-\mathcal{I}[t]$. By labeling $F_w[t]=0$ for any $w\in \mathcal{E}[t]$, UAV-$u$ observes all the WNs' actual node types. For the observation of $\{B_1^u [t],...,B_W^u [t]\}$ and $\{C_1^u [t],...,C_W^u [t]\}$, while the I-nodes' actual battery energy levels and accumulated data size are obtained based on their reporting, the E-nodes' battery energy levels and accumulated data size are generally partially observable at UAV-$u$ as given in (\ref{equ: uav knows battery}) and (\ref{equ: uav knows data}), respectively. 
	
\begin{figure}[h]
		\setlength{\abovecaptionskip}{-0.0in}
		\centering
		\DeclareGraphicsExtensions{.eps,.mps,.pdf,.jpg,.png}
		\DeclareGraphicsRule{*}{eps}{*}{}
		\includegraphics[angle=0, width=0.49 \textwidth]{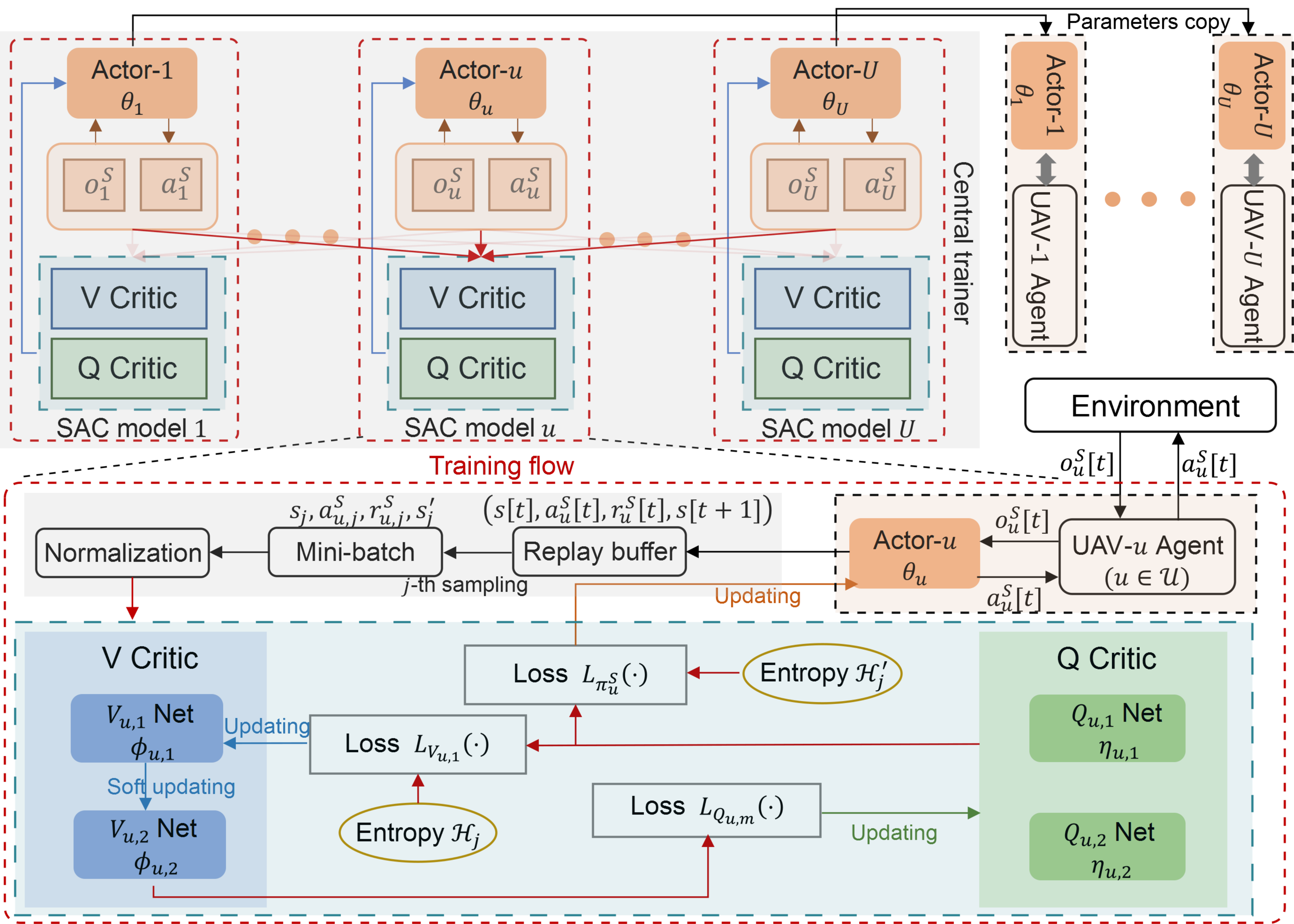}
		\caption{Central training of the SAC policy.}
		\label{fig: SAC Training}
		
	\end{figure}

	As a result, since $U<W$ generally holds in practice, the gathered observations from all the $U$ UAVs at the central trainer only provides partial information about all the $W$ WNs' status in slot $t$ in general. This leads to a POMDP modeling at the central trainer. Denote the POMDP state \cite{markov} in slot $t$ as $s[t]=\{o_1^S[t],...,o_U^S[t]\}$, which is stored in the replay buffer at the central trainer, as shown in Fig.~\ref{fig: SAC Training}. The POMDP state space is denoted as $\mathcal{S}$ with $s[t] \in \mathcal{S}$.

	\subsection{Action and Reward Function} \label{subsection: SAC action and reward}
	For each UAV-$u$, the output of the SAC policy in each slot $t$ contains UAV-$u$'s horizontal moving angle $\varphi_u[t]$, velocity $V_u[t]$ and WET decision $Z_u[t]$, where $q_u[t]$ is easily obtained from $(\varphi_u[t],V_u[t])$. We thus denote the action of UAV-$u$ in the POMDP model as $a_u^S [t]=\{\varphi_u[t],V_u [t],Z_u [t]\}$, where ($\varphi_u[t],V_u[t]$) is assured to satisfy the constraint in (\ref{constraint: b}) by using the range mapping function in \cite{oubbati2022synchronizing}. Since the SAC has continuous action output \cite{sac}, to satisfy the binary WET decision constraint in (\ref{constraint: d}), we let $Z_u [t]=1$ when the SAC policy network's WET decision output is positive, or $Z_u [t]=0$, otherwise. The action space is denoted as $\mathcal{A}^S$ with $a_u^S[t]\in \mathcal{A}^S$.
	
	In each slot $t$, based on UAV-$u$'s observation $o_u^S[t]$, it implements an action $a_u^S[t]\in \mathcal{A}^S$, and obtains a reward at the end of slot $t$. As will be specified in the next subsection, the trajectory determined by the SAC policy is used as a known state for the POMDP modeling of the DQN for WDC decisions. Hence, to find a proper trajectory for each UAV's WET in tier-1 and WDC in tier-2, the SAC reward function $r_u^S[t]$ is designed to achieve a proper trade-off between maximizing the E-nodes' harvested energy and the I-nodes' transmission data size under the constraints given in (\ref{constraint: a}), (\ref{constraint: c}) and (\ref{constraint: e}) for problem (P1). To be specific, the reward function includes the following four parts:
	
	\emph{1) Reward on WET to E-nodes:} Due to the generally different battery energy of the E-nodes, their energy demands are different in general. To provide on-demand WET to the E-nodes, we adopt the metric of HoE to measure each WN's time-varying energy demands as in \cite{magrl}. Denote $H_w[t]$ as the HoE of WN-$w$ in slot $t$. If $w\in \mathcal{I}[t]$, WN-$w$ is an I-node with $H_w[t]=0$; and If $w\in \mathcal{E}[t]$, WN-$w$ is an E-node with its $H_w[t]$ given as:
	\small
	\begin{equation}
		\vspace{-0.00in}
		\label{equ: wireless node HoE update}
		H_w[t]\!= \! \begin{cases}
			\! H_w[t \! - \! 1] \! + \! 1,  &if ~ E_w^{har}[t \! - \! 1] \!< \! E^{exp}, \\
			\max \! \left( H_w[t \!-\!1]\!-\!1,\! 1\right),  &if~ E_w^{har}[t \! - \! 1] \! \ge \! E^{exp},
		\end{cases}
	\end{equation}
	\normalsize
	where $E^{exp}=\frac{B_I-B_E}{T}$ represents the average energy amount that is expected to harvest at each E-node in each slot, such that it can transform to an I-node based on (\ref{equ: wireless node flag update}) for at least one time on average during the $T$ slots. From (\ref{equ: wireless node HoE update}), the HoE of E-node-$w$ is increased by $1$ if $E_w^{har}[t \! - \! 1] \!< \! E^{exp}$, or reduced by 1, otherwise. In each slot $t$, the minimum HoE is set to $1$ for all the E-nodes. Denote UAV-$u$'s reward on WET to all the E-nodes in slot $t$ by $b_u^{WET}[t]$ and is given as 
	\small
	\begin{equation} 
		\vspace{-0.00in}
		\label{equ: reward for charge}
		b_u^{WET}[t] = N_u[t] \cdot \sum_{w\in \mathcal{E}[t]}H_w[t] (B_w^u[t+1]-B_w^u[t]),
	\end{equation}
	\normalsize
	where $N_u[t] \!=\! \sum_{w\in \mathcal{E}[t]}\frac{\bar{P}(P_UZ_u[t]G_w^u[t])\vartheta}{B_w^u[t+1]-B_w^u[t]}$ is UAV-$u$'s WET weight by charging all the E-nodes, with $\frac{\bar{P}(P_UZ_u[t]G_w^u[t])\vartheta}{B_w^u[t+1]-B_w^u[t]}$ representing the ratio of E-node-$w$'s harvested DC energy from UAV-$u$ to that from all the UAVs, and $\sum_{w\in \mathcal{E}[t]}H_w[t] (B_w^u[t+1]-B_w^u[t])$ is the sum of each E-node's increased battery energy in slot $t$ biased by its HoE. If $B_w^u[t+1]-B_w^u[t]$ is zero for a particular E-node-$w$, since $\bar{P}(P_UZ_u[t]G_w^u[t])\vartheta=0$ is also obtained, as detailed in Section \ref{subsection: sac pomdp}, we set $\frac{\bar{P}(P_UZ_u[t]G_w^u[t])\vartheta}{B_w^u[t+1]-B_w^u[t]}=0$. It is easy to find from (\ref{equ: reward for charge}) that UAV-$u$ generally gets more reward by transmitting energy to the E-nodes with higher HoE and/or higher $G_w^u[t]$. Hence, $b_u^{WET}[t]$ reflects the effects of UAV-$u$'s WET on increasing all the E-nodes' battery energy according to their demands in slot $t$.
	
	2) \emph{Reward on WDC from I-nodes:} Let $b_u^{WDC}[t]$ represent UAV-$u$'s reward on collecting data from the I-nodes in slot $t$, which is expressed as
	\small
	\begin{equation}
		\vspace{-0.00in}
		\label{equ: reward for information}
		b_u^{WDC}[t] = \sum_{w\in \mathcal{I}[t]}\left( \sum_{t^{'}=1}^{t}C_w[t^{'}] - \frac{t}{T}C_{min} \right).
	\end{equation}
	\normalsize
	In (\ref{equ: reward for information}), $\sum_{t^{'}=1}^{t}C_w[t^{'}]$ or $\frac{t}{T}C_{min}$ gives I-node-$w$'s accumulated data size that is already transmitted or expected to transmit on average from the first slot to the current slot $t$, respectively. The larger value $b_u^{WDC}[t]$ achieves, the larger reward UAV-$u$ obtains to meet the constraint in (\ref{constraint: a}). It is noted that under the UAV's trajectory from the SAC policy, $\sum_{t^{'}=1}^{t}C_w[t^{'}]$ is determined by the UAV's WDC decisions from the DQN policy in tier-2. Hence, the SAC policy and the DQN policy in the two tiers are mutually affected in general.
	
	3) \emph{Reward on Energy Saving}: Denote $b_u^{ES}[t]$ as UAV-$u$'s reward on saving its own battery energy in slot $t$ to meet the constraint in (\ref{constraint: c}). We have $b_u^{ES}[t] = B_u[t+1]-B_U^{min}$, which requires to be positive for maximizing UAV-$u$'s reward.
	
	4) \emph{Reward on Safe Distance}: Denote $b_u^{SD}[t]$ as UAV-$u$'s reward on keeping the safe distance of $d_{min}$ to any other UAV to meet the constraint in (\ref{constraint: e}), where we set $b_u^{SD}[t]=-1$ if the constraint in (\ref{constraint: e}) is not satisfied, or $b_u^{SD}[t]=0$, otherwise.
	
	Since all the four rewards have different units and magnitudes, they cannot be directly summed up. We multiply the non-negative coefficients $\xi_1$, $\xi_2$, $\xi_3$ and $\xi_4$ to $b_u^{WET}[t]$, $b_u^{WDC}[t]$, $b_u^{ES}[t]$ and $b_u^{SD}[t]$, respectively, to assure the different rewards have the same or similar magnitudes, and obtain the overall reward function as $r_u^S[t]=\xi_1 b_u^{WET}[t] + \xi_2 b_u^{WDC}[t] + \xi_3 b_u^{ES}[t] + \xi_4 b_u^{SD}[t]$. The values of the $4$ coefficients for simulations are given in Table \ref{table: parameters set}. We denote the reward space as $\mathcal{R}^S$ with $r_u^S[t] \in \mathcal{R}^S$.
	
	\vspace{-0.1in}

	\subsection{SAC Training at the Central Trainer} \label{subsection: sac training}
	
	Given the limited computational resources available to each UAV and the restricted range for the E-nodes' status reporting, the SAC policy is centrally trained at the central trainer, to enhance the POMDP state observation completeness and alleviate the computational burden on the UAVs. As shown in Fig.~\ref{fig: SAC Training}, there are a total of $U$ parallelly-trained SAC models at the central trainer, each for one UAV. Each SAC model consists of five neural networks, which are the policy network Actor-$u$ with parameter $\theta_u$, the two state networks $V_{u,1}$ Net and $V_{u,2}$ Net with parameters $\phi_{u,1}$ and $\phi_{u,2}$, respectively, for V critic, and the two state-action networks $Q_{u,1}$ Net and $Q_{u,2}$ Net with parameters $\eta_{u,1}$ and $\eta_{u,2}$, respectively, for Q critic. After the central training, the well-trained parameters of the  Actor-$u$ network at the central trainer is then copied to UAV-$u$.
	
	From Fig.~\ref{fig: SAC Training}, the two V critic networks and the two Q critic networks, are used to assist the training of Actor-$u$, and the entropy $\mathcal{H}(\cdot)$ that is used to design the loss function can enhance the agent UAV-$u$'s exploration of the environment \cite{sac}. To be specific, the policy network Actor-$u$ trains the policy function $\pi_u^S(\cdot)$ that maps the observation $o_u^S [t]$ of UAV-$u$ to its action $a_u^S [t]$, $V_{u,1}$ Net and $V_{u,2}$ Net train the state-action functions $V_{u,1}(\cdot)$ and $V_{u,2}(\cdot)$, respectively, and $Q_{u,1}$ Net and $Q_{u,2}$ Net train the state functions $Q_{u,1}(\cdot)$ and $Q_{u,2}(\cdot)$, respectively. The state functions $V_{u,1}(\cdot)$ and $V_{u,2}(\cdot)$ estimate the value of the cumulative discount reward received by UAV-$u$ in state $s[t]$. We have $V_{u,1}(s[t]) \!=\! \mathbb{E}_{a_u^{'} \sim \pi_u^S} \left[Q_{min} \left(s[t],a_u^S [t] \right) \!-\! \alpha_u \mathcal{H}(\pi_u^S ) \right]$, where $\mathcal{H}(\pi_u^S ) \!=\! \log \left( \pi_{u}^S (a_u^{'}|o_u^S[t]) \right)$ is the entropy of the output action $a_u^{'}$ based on policy $\pi_u^S$ in observation $o_u^S[t]$, $a_u^{'} \! \sim \! \pi_u^S$ denotes the action $a_u^{'}$ taken from the policy $\pi_u^S$, $Q_{min}(\cdot)\!=\! \min \! \left( Q_{u,1}(\cdot), Q_{u,2}(\cdot) \right)$, and $\alpha_u$ is the temperature coefficient (i.e., the entropy weight). Generally, $\alpha_u$ should be reduced with the number of trainings in order to ensure the convergence of Actor-$u$. Moreover, the state-action functions $Q_{u,1}(\cdot)$ and $Q_{u,2}(\cdot)$ estimate the cumulative discount reward value when UAV-$u$ executes the action $a_u^S [t]$ in state $s[t]$. We have $Q_{u,m}(s[t],a_u^S[t])\! =\! r_u^S[t] \! + \! \gamma \mathbb{E}[V_{u,2}(s[t+1])]$, where $\gamma$ is the discount factor and $\mathbb{E}[\cdot]$ is the operation of expectation. The goal is to find the optimal policy $\pi_u^{S*} \! = \! \arg \max_{\pi_u^S} \sum_{t\in\mathcal{T}} \mathbb{E}[r_u^S[t]+\alpha_u \mathcal{H}(\pi_u^S)]$ that maximizes the reward and the entropy gained at UAV-$u$.
	
	The parameter $\theta_u$ of the Actor-$u$ network is sent from the central trainer to each UAV-$u$ at regular intervals, and each UAV-$u$ copies the received parameter $\theta_u$ to its local Actor-$u$ network, which outputs UAV-$u$'s action $a_u^S[t]\in \mathcal{A}^S$ based on its local observation $o_u^S[t]\in \mathcal{O}^S$ in each slot $t$. After implementing the action $a_u^S[t]$, UAV-$u$ obtains the reward $r_u^S[t]\in \mathcal{R}^S$ and the new observation $o_u^S[t+1]$ at the beginning of slot $t+1$. As shown in Fig.~\ref{fig: system model}, the central trainer receives the combined information ($s[t]$, $a_u^S [t]$, $r_u^S [t]$, $s[t+1]$) from each UAV with $s[t]=\{o_1^S[t],...,o_U^S[t] \}$ as detailed in Section \ref{subsection: sac pomdp}, and stores it in the replay buffer. 
	
	The parameters of the five neural networks in the SAC model $u$ are updated to minimize the corresponding loss functions shown in Fig.~\ref{fig: SAC Training}. Specifically, the $j$-th training of the SAC model $u$ is based on the experience of taking mini-batch ($s_j$, $a_{u,j}^S$, $r_{u,j}^S$, $s_j^{'}$) from the replay buffer. Denote $\phi_{u,1}$ as the parameter for the $V_{u,1}$ Net. UAV-$u$ uses the loss function
		\small
		\begin{equation}
			\vspace{-0.0in}
			\label{equ: SAC Loss V}
			L_{V_{u,1}}(\phi_{u,1}) \! =\! \mathbb{E}\left [\! \frac{1}{2} \! \left( \!  V_{u,1}(s_{j})\!-\!
			\mathbb{E}_{a_{u}^{'}\sim\pi_{u}^S}\left [  \!  Q_{min}(s_{j},a_u^{'}) \! -  \! 
			\alpha_u \mathcal{H}_j  \!  \right ]   \!
			\right)^2  \!  \right ],
		\end{equation}
		\normalsize
		where the entropy $\mathcal{H}_j = \log \left(\pi_{u}^S(a_u^{'}|o_{u,j}^S) \right)$. For the parameter $\phi_{u,2}$ of the $V_{u,2}$ Net, we apply a soft update with $\phi_{u,2} \! \gets \! \tau \phi_{u,2} \! + \! (1 \!- \! \tau) \phi_{u,1}$, $\tau\in [0,1)$. Moreover, denote $\eta_{u,m}$ as the parameter for the $Q_{u,m}$ Net, $\forall m\in \{1,2\}$. The loss function for the $Q_{u,m}$ Net is
		\small
		\begin{equation}
			\vspace{-0.0in}
			\label{equ: SAC Loss Q}
			L_{Q_{u,m}} (\eta_{u,m}) \! = \! \mathbb{E}\left [ \!  \frac{1}{2}\left ( \! 
			Q_{u,m}(s_{j},a_{u,j}^S) \! - \! \left(r_{u,j}^S+\gamma\mathbb{E}\left [ \! V_{u,2}(s_{j}^{'}) \!  \right ]  \! \right )  \right )^2   \!  \right ].
		\end{equation}
		\normalsize
		For the SAC policy network Actor-$u$, we apply the following loss function: 
		\small
		\begin{equation}
			\vspace{-0.0in}
			\label{equ: SAC Loss Policy 2}
			L_{\pi_{u}^S}(\theta_u)=\mathbb{E}_{\varepsilon \sim\chi}\left [ 
			\alpha_u \mathcal{H}_j^{'}-Q_{min}(s_{j},a_{\theta_u}(o_{u,j}^S;\varepsilon))
			\right],  
		\end{equation}
		\normalsize
		where unlike the entropy $\mathcal{H}_j$ in (\ref{equ: SAC Loss V}),  the entropy $\mathcal{H}_j^{'} \!=\! \log \left( \pi_u^S \left( a_{\theta_u} (o_{u,j}^S; \varepsilon) | o_{u,j}^S \right) \right)$ is calculated from the noise-action $a_{\theta_u} (o_{u,j}^S; \varepsilon)$, where $\varepsilon \sim \chi$ is the noise sample taken from a fixed distribution $\chi$. According to \cite{sac}, adding noise to the actions during the training can avoid network overfitting, and at the same time, increase UAV-$u$'s exploration of the environment.
	At last, for the update of the temperature coefficient $\alpha_u$, we use the loss function
	\small
	\begin{equation}
		\vspace{-0.0in}
		\label{loss: tempture}
		L_{\alpha_u}(\alpha_u) = \mathbb{E}_{a_u^{'}\sim \pi_{u}^S} \left[ -\alpha_u \log \left( \pi_u^S(a_u^{'}|o_{u,j}^S) \right)-\alpha_u \tilde{H} \right],
	\end{equation}
	\normalsize
	where $\tilde{H}$ is a constant that is generally equal to the dimension of the observation $o_u^S$ (i.e., $\tilde{H} \! = \! |o_u^S|$). 
	All the parameters $\phi_{u,1}$, $\eta_{u,m}$, $\theta_u$, $\alpha_u$ are updated as the optimal values that minimize the corresponding loss functions given in (\ref{equ: SAC Loss V})-(\ref{loss: tempture}), respectively, by using the Adaptive moment estimation (Adam) optimization algorithm \cite{adam}.

	\section{DQN in Tier-2} \label{section: DQN}

	\subsection{POMDP Modeling for WDC Decisions} \label{section: DQN pomdp}
	
	Based on the determined trajectory from the SAC in tier-1, DQN in tier-2 is used to determine each UAV's binary WDC decisions over the $K$ sub-slots. As compared to the SAC training, since the DQN model is simpler, to avoid large transmission overhead of neural network parameters with the central trainer, the training of the DQN in tier-2 is performed at each of the UAVs locally.
	
	\subsubsection{Observation}
	Similar to the case for observing the WNs' status in Section \ref{subsection: sac pomdp}, each UAV obtains its observation via the WNs' periodic status reporting. Here,  considering the trivially-changed environment over the sub-slots in the same slot, instead of impelling each WN to additionally report its status over sub-slots, we only use the WNs' reported status at the beginning of each slot $t$. Denote, the observation of UAV-$u$ at sub-slot $k$ in slot $t$ is $o_u^{D,t} [k]=\{x_u[t],y_u[t],C_1^u [t],...,C_W^u [t],k\}$, which consists of UAV-$u$'s own horizontal location in slot $t$, all the WNs' accumulated transmission data size from $t^{'} \!=\! 1$ to the previous slot $t \!-\!1$, and the sub-slot index $k$. In $o_u^{D,t}[k]$, $(x_u[t],y_u[t])$ is determined by the SAC policy in tier-1, $\{C_w^u[t]\}$ only provide partial information on all the WNs' accumulated transmission data as specified in Section \ref{subsection: sac pomdp}, and $k$ is used to distinguish different sub-slots. By using $o_u^{D,t}[k]$ as the observation for the DQN, we also obtain a POMDP for the UAV-$u$'s WDC decisions. Denote $\mathcal{O}^D$ as the observation space with $o_u^{D,t}[k] \in \mathcal{O}^D$.
	
	\vspace{0.02in}
	\subsubsection{Action and Reward} \label{subsubsection: DQN Action and Reward}
	Denote $a_u^{D,t}[k] \!=\! w \! \in \! \mathcal{W}$ as the POMDP action of UAV-$u$ at sub-slot $k$ in slot $t$. Given an observation $o_u^{D,t}[k]\in \mathcal{O}^D$, the DQN's output is a Q-vector $Q(o_u^{D,t}[k])$ of length $W$, where the $w$-th element gives the probability of scheduling WN-$w\in \mathcal{W}$. To satisfy the one-to-one user association constraint in (\ref{constraint: scheduling}), we consider the following steps to determine each $D_{u,w}^t[k]$: a) UAV-$u$ initializes a WN candidate set $\mathcal{W}_{ca}=\mathcal{W}$; b) UAV-$u$ sets $D_{u,w}^t[k]=1$ for $w=\arg\max_{w\in \mathcal{W}_{ca}}Q(o_u^{D,t}[k])$, and sets $D_{u,i}^t[k]=0$, $\forall i \in \mathcal{W}, i\neq w$; c) if the selected WN-$w$ is an I-node, UAV-$u$ sends an association request to WN-$w$; if WN-$w$ is an E-node or WN-$w$ feeds back UAV-$u$ that UAV-$u^{'}$ with $u^{'}<u$ also selects it, UAV-$u$ sets $D_{u,w}[k]=0$, updates $\mathcal{W}_{ca}=\mathcal{W}_{ca}-\{w\}$, and goes to step b). The above steps repeat until the selected I-node WN-$w$ feeds back an association confirmation to UAV-$u$, or $\mathcal{W}_{ca}=\emptyset$ is obtained. For the former case, UAV-$u$ finds a proper I-node-$w$ as its POMDP action that meets the constraint in (\ref{constraint: scheduling}), and for the latter case, UAV-$u$ remains silent. Since $U<W$ and the required one-to-one association in (\ref{constraint: scheduling}), the above procedure is rapid to find $a_u^{D,t}[k]$. Denote the DQN action space as $\mathcal{A}^D$ with $a_u^{D,t}[k] \in \mathcal{A}^D$. For UAV-$u\in \mathcal{U}$, the reward function is expressed as
	
	\small
	\begin{equation}
		\vspace{-0.0in}
		\label{equ: dqn reward}
		r_u^{D,t}[k] \!=\! \sum_{w\in \mathcal{I}[t]} M_{u,w}^{t}[k] \!+\! \sum_{w \in \mathcal{I}[t]}\left( C_w^u[t] \!-\! \frac{t \!-\! 1}{T}C_{min} \right).
	\end{equation}
	\normalsize
	In (\ref{equ: dqn reward}), the first item $\sum_{w\in \mathcal{I}[t]} M_{u,w}^{t}[k]$ is UAV-$u$'s received data size at sub-slot $k$ in slot $t$ under the constraint in (\ref{constraint: scheduling}), and the second item \small $\sum_{w \in \mathcal{I}[t]}\left( C_w^u[t] \!-\! \frac{t \!-\! 1}{T}C_{min} \right)$\normalsize is the sum of all I-nodes' gap between their accumulated data size and the expected transmission data size $\frac{t \!-\! 1}{T}C_{min}$ over the past $t \!-\! 1$ slots. With proper WDC decisions, the UAV obtains large $\sum_{w\in \mathcal{I}[t]} M_{u,w}^{t}[k]$ to assure the constraint in (\ref{constraint: a}).

	\subsection{DQN Training at each local UAV} \label{subsection: dqn training}

\begin{figure}[h]
		\setlength{\abovecaptionskip}{-0.0in}
		\centering
		\DeclareGraphicsExtensions{.eps,.mps,.pdf,.jpg,.png}
		\DeclareGraphicsRule{*}{eps}{*}{}
		\includegraphics[angle=0, width=0.45 \textwidth]{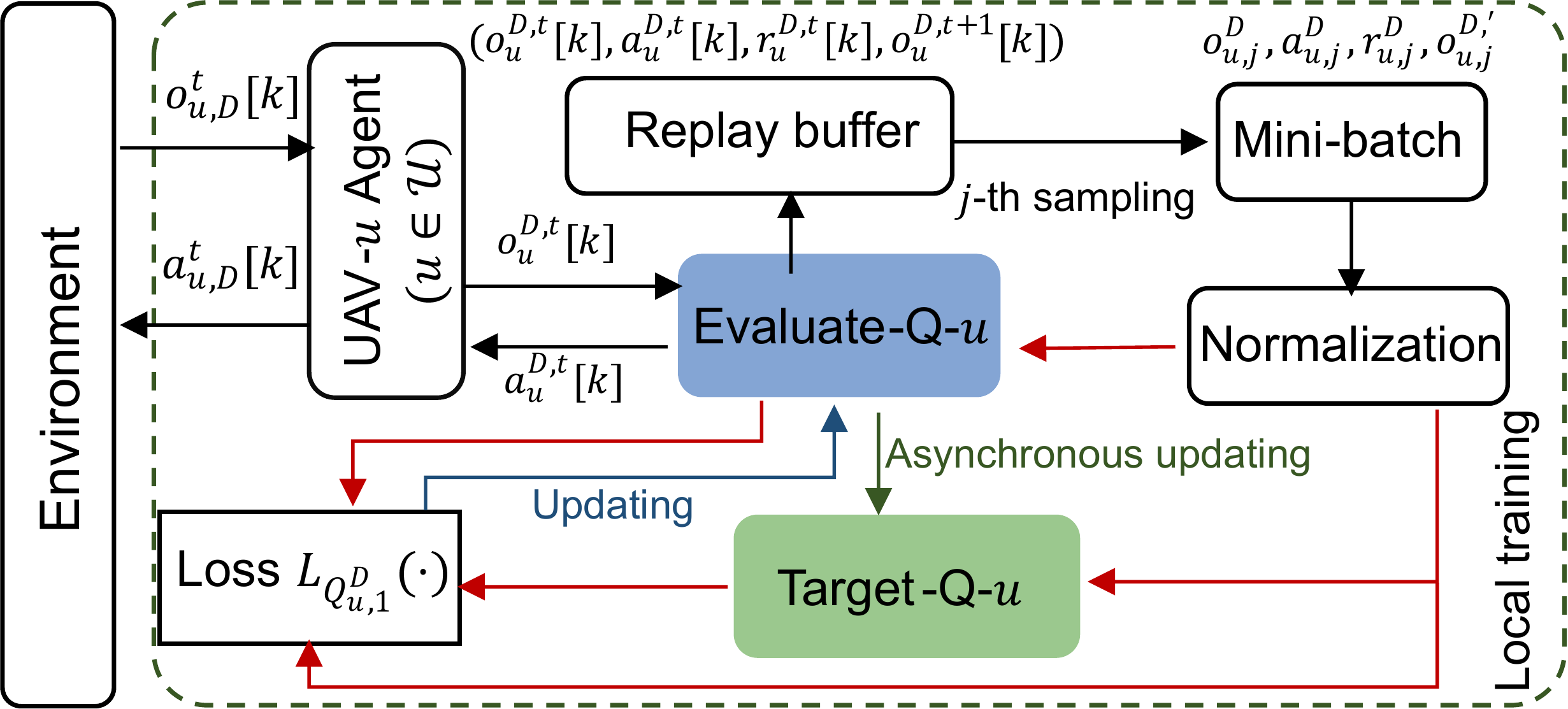}
		\caption{Local training of the DQN at UAV-$u$.}
		\label{fig: DQN Training}
		
	\end{figure}

	As shown in Fig.~\ref{fig: DQN Training}, the DQN model at each local UAV contains two neural networks, which are the Evaluate-Q-$u$ network and the Target-Q-$u$ network with parameter $\lambda_{u,1}$, $\lambda_{u,2}$, respectively. The Evaluate-Q-$u$ network is trained as the state-action function $Q_{u,1}^D(\cdot)$, which gives the UAV-$u$'s Q-value (i.e., the cumulative discount reward) distribution for different actions under the observation $o_u^{D,t}[k]$. The Target-Q-$u$ network is trained as a state-action function $Q_{u,2}^D(\cdot)$ to guide the training of the Evaluate-Q-$u$ network. The goal is to enable UAV-$u$ to maximize its own cumulative discount reward $Q_{u,1}^{D*}(o,a)$. According to \cite{dqn}, we have $Q_{u,1}^{D*}(o,a)=r+\gamma \sum_{o^{'}\in \mathcal{O}_D}\mathcal{P}_{o,o^{'}}^a \max_{a^{'}}Q_{u,1}^{D*}(o^{'},a^{'})$, where $o=o_u^{D,t}[k]$, $a=a_u^{D,t}[k]$, $r=r_u^{D,t}[k]$, $o^{'}=o_u^{D,t+1}[k]$, $a^{'}$ is the action with maximum Q-value at $o^{'}$, and $\mathcal{P}_{o,o^{'}}^a$ is the probability that UAV-$u$ executes action $a$ to get the next observation $o^{'}$ at the current observation $o$.
	
	The parameter $\lambda_{u,1}$ of the Evaluate-Q-$u$ network is updated according to the following loss function
	\small
	\begin{equation}
		\vspace{-0.0in}
		\label{loss: Q}
		L_{Q_{u,1}^D}(\lambda_{u,1})=\mathbb{E}\left[\frac{1}{2}\left(Q_{u,1}^D(o_{u,j}^D,a_{u,j}^D)-Q_{target}\right)^2\right],
	\end{equation}
	\normalsize
	where $Q_{target}=\left(r_{u,j}^D+\gamma\max_{a^{'}}Q_{u,2}^D\left( o_{u,j}^{D,'},a^{'}\right)\right)$, and the $j$-th training is performed based on the experience of taking a mini-batch ($o_{u,j}^D$, $a_{u,j}^D$, $r_{u,j}^D$, $o_{u,j}^{D,'}$) from the local replay buffer. For the Target-Q-$u$ network with parameter $\lambda_{u,2}$, we perform the asynchronous update based on the parameter $\lambda_{u,1}$, similar to the process in \cite{dqn}.
	
	\subsection{MAHDRL Algorithm} \label{section: MAHDRL}
	Based on the SAC model in Section \ref{section: SAC} and the DQN model in Section \ref{section: DQN}, the MAHDRL algorithm to solve the problem (P1) is given in \textbf{Algorithm 1}. In summary, for the SAC training in tier-1, the maximization of the WET reward in (\ref{equ: reward for charge}) guides the UAV to learn along the direction of increasing each E-node's battery energy so as to transform more E-nodes into the I-nodes and, at the same time, the maximization of the WDC reward in (\ref{equ: reward for information}) leads each UAV to increase $C_{total}$ from the transformed I-nodes when determining its trajectory, while the reward maximizations of $b_u^{ES}[t]$ and $b_u^{SD}[t]$ guide the UAVs to meet the constraints in (\ref{constraint: c}) and (\ref{constraint: e}), respectively. In tier-2, the maximization of the DQN reward in (\ref{equ: dqn reward}) further leads each UAV to maximize $C_{total}$ under the constraint in (\ref{constraint: a}). The constraints in (\ref{constraint: scheduling}), (\ref{constraint: b}) and (\ref{constraint: d}) are properly satisfied by the action output of the SAC and the DQN policies. Therefore, the proposed MAHDRL approach can properly solve problem (P1).
	
	\begin{algorithm}
		\small
		\caption{MAHDRL Algorithm}
		\begin{algorithmic}[1]
			\STATE {\textbf{Initialize} replay buffer, learning rate, discount factor $\gamma$, soft update weight $\tau$ and temperature factor $\alpha_u$, $u\in\mathcal{U}$. Initialize the parameters of SAC's all five networks and DQN's two networks;}
			\FOR{Episode $\gets1,...,EPS$}
			\STATE{Initialize the locations and battery levels of all UAVs and WNs;}
			\STATE{Initialize the observation $o_{u}^S[0]$, $o_u^{D,0}[1]$, ..., $o_u^{D,0}[K]$ and SAC state $s[0]$, $\forall u\in \mathcal{U}$;}
			\FOR{$t\gets 1,..., T$}
			\STATE{\textbf{get action} $a_u^S[t]$ and $a_u^{D,t}[k]$ based on Sections \ref{subsection: SAC action and reward} and \ref{subsubsection: DQN Action and Reward}, respectively;}
			\STATE{\textbf{execute action} $a_{u}^S[t]$, and $a_u^{D,t}[k]$, based on Sections \ref{subsection: SAC action and reward} and \ref{subsubsection: DQN Action and Reward}, respectively;}
			\STATE{\textbf{store the experience} $\left(s[t],a_{u}^S[t],r_u^S[t],s[t \!+\!1]\right)$ and $\left(o_u^{D,t}[k],a_u^{D,t}[k],r_u^{D,t}[k],o_u^{D,t+1}[k]\right)$ into the replay buffer;}
			\STATE{using the loss functions in (\ref{equ: SAC Loss V}), (\ref{equ: SAC Loss Q}), (\ref{equ: SAC Loss Policy 2}), (\ref{loss: tempture}) and (\ref{loss: Q}) to \textbf{update} the SAC's parameters based on Section \ref{subsection: sac training} and \textbf{update} the DQN's parameters based on Section \ref{subsection: dqn training}, respectively;}
			\STATE{$o_u^S[t]\gets o_u^S[t+1]$, $o_u^{D,t}[k]\gets o_u^{D,t+1}[k]$ and $s[t]\gets s[t+1]$;}
			\ENDFOR
			\ENDFOR
		\end{algorithmic}
		\normalsize
	\end{algorithm}
	
	We now analyze the time complexity of implementing \textbf{Algorithm 1}. Let $L_A=5$ and $N_{A,j}$, $L_V=5$ and $N_{V,j}$, and $L_Q=5$ and $N_{Q,j}$ represent the number of fully connected layers and the number of neurons in the $j$-th layer of Actor-$u$ network, $V_{u,m}$ ($m\in\{1,2\}$) Net, and $Q_{u,m}$ Net, respectively. The time complexity of training the SAC at one step according to \cite{sac} is \small$\mathcal{O}\left(\sum_{j = \!2}^{L_A}N_{A,j \!-\! 1} \!\cdot \! N_{A,j} \!+\! \sum_{j = \!2}^{L_V}N_{V,j \!-\! 1} \!\cdot \! N_{V,j} \!+ \! \sum_{j = 2}^{L_Q}N_{Q,j \!-\! 1} \!\cdot \! N_{Q,j} \right)$\normalsize. Denote $L_E=5$ and $N_{E,j}$ as the number of fully connected layers and the number of neurons in the $j$-th layer of the Evaluate-Q-$u$ network, $u\in \mathcal{U}$, respectively. The time complexity of training the DQN at one step according to \cite{dqn} is \small $\mathcal{O}\left( \sum_{j = 2}^{L_E}N_{E,j \!-\!1}\cdot N_{E,j} \right)$\normalsize. Therefore, the time complexity of the proposed MAHDRL algorithm is \small $\mathcal{O}\left(U \! \times \! EPS \! \times \! T \! \times \! \left(\sum_{j = 2}^{L_E}N_{E,j \!-\! 1} \! \cdot \! N_{E,j} \! + \! \sum_{j = 2}^{L_A}N_{A,j \!-\! 1} \! \cdot \! N_{A,j}\right.\right.$ $\left. \left. +\sum_{j = 2}^{L_V}N_{V,j \!-\! 1}\cdot N_{V,j}+\sum_{j = 2}^{L_Q}N_{Q, j \!-\! 1} \!\cdot\! N_{Q,j} \right) \right)$\normalsize, which increases linearly over the number of the UAVs.

	\section{Simulation Results} \label{section: simulation}
	Simulation results are provided in this section to validate the performance of the proposed MAHDRL approach in \textbf{Algorithm 1}. As specified in Sections \ref{subsection: sac training} and \ref{subsection: dqn training}, in the training stage, the SAC policy is first trained centrally at the central trainer, where the well-trained parameters are then copied to tier-1 in each of the UAV agent's DRL model. The DQN policy in tier-2 is trained locally at each of the UAV agent. In the test stage, all the UAVs distributively make their WET, WDC and trajectory decisions based on their own policy networks.
	
	We perform simulations based on python-3.9.12 and pytorch-1.12.1. Unless otherwise stated, in all the simulations, we consider an area of $400$ m$\times 400$ m and set $U=4$ UAVs with random start locations, $W=10$ WNs with random initial battery energy in the range of $[2, 4]$ mW$\cdot$s, the mission period $T \!=\! 300$ s, $K \!=\! 4$ sub-slots, and E-nodes' reporting range $d_{\textrm{cov}} \!=\! 20$ m. The UAVs' propulsion power model parameters follow that in \cite{zeng2019energy}. For the neural networks of the SAC and the DQN, we set the input layers with $N_{A,1} \!=\! 3W \!+\! 3$, $N_{V,1}=(3W \!+\! 3)\! \times \! U$, $N_{Q,1}=(3W \!+\! 3)\! \times \! U \!+\! 3$ and $N_{E,1} \!=\! W \!+\! 2$, respectively, and set the output layers with $N_{A,5} \!=\! 3$, $N_{V,5} \!=\! 1$, $N_{Q,5} \!=\! 1$ and $N_{E,5} \!=\! W$, respectively. According to \cite{wang2021deep,wang2019multi} and \cite{barto2021}, for all the neural networks, $256$ neurons are set in the hidden layers. For the SAC in tier-1, we set the learning rates for all the Actor, the Q Critic, and the V Critic networks to $0.0003$, and set the temperature coefficient $\alpha_u$'s learning rate to $0.0002$. For the DQN at tier-2, its learning rate is set to decay from $0.01$ to $0.000001$, and its exploration rate of the environment is set to decay from $0.9$ to $0.02$. Other simulation parameters are given in TABLE \ref{table: parameters set}.
	
	\begin{table}[h]
		\centering
		\caption{ Simulation parameters}\label{table: parameters set}
		\vspace{-1em}
		\scriptsize
		\setlength{\tabcolsep}{1  mm}{
			\begin{tabular}{|c|c||c|c|}
				\hline
				\textbf{Parameter}   & \textbf{Value}   & \textbf{Parameter}   & \textbf{Value}      \\ \hline
				$h$       &  $5$ m  &  $\sigma^2$  & $-90$ dBm \\ \hline
				$P_U$, $P_W$, $P_I$  & $1$ W, $0.1$ mW, $10$ mW  &  $P_{sen}, P_{sat}$  & $-10$, $7$ dBm \\ \hline
				$\alpha_L, \alpha_N$       &  $3$, $5$  &  $B_E$, $B_I$  & $2$, $4$ mW$\cdot$s \\ \hline
				$B_u^{min}$, $B_u^{max}$      &  $3e4$, $4e5$ W$\cdot$s &  $d_{min}$, $C_{min}$   &   $2$ m, $100$ bits/Hz  \\ \hline
				$\epsilon$-greedy strategy  & $0.01\to 1e \!-\! 6$ & \small$\xi_1,\xi_2,\xi_3,\xi_4$  & $20$, $0.01$, $1e \!-\! 6$, $1 \!$ \normalsize \\ \hline
				$\gamma,\tau$  &  $0.99$, $0.999$        &  $a,b$ for $P_{LoS}$   & $12.08$, $0.11$   \\ \hline
				replay buffer's size &  $131072$   &  mini-batch's size   & $128$ \\ \hline
		\end{tabular}}
	\end{table}

	\begin{figure}[htbp]
		\setlength{\abovecaptionskip}{-0.0in}
		\centering
		\DeclareGraphicsExtensions{.eps,.mps,.pdf,.jpg,.png}
		\DeclareGraphicsRule{*}{eps}{*}{}
		\includegraphics[angle=0, width=0.45 \textwidth]{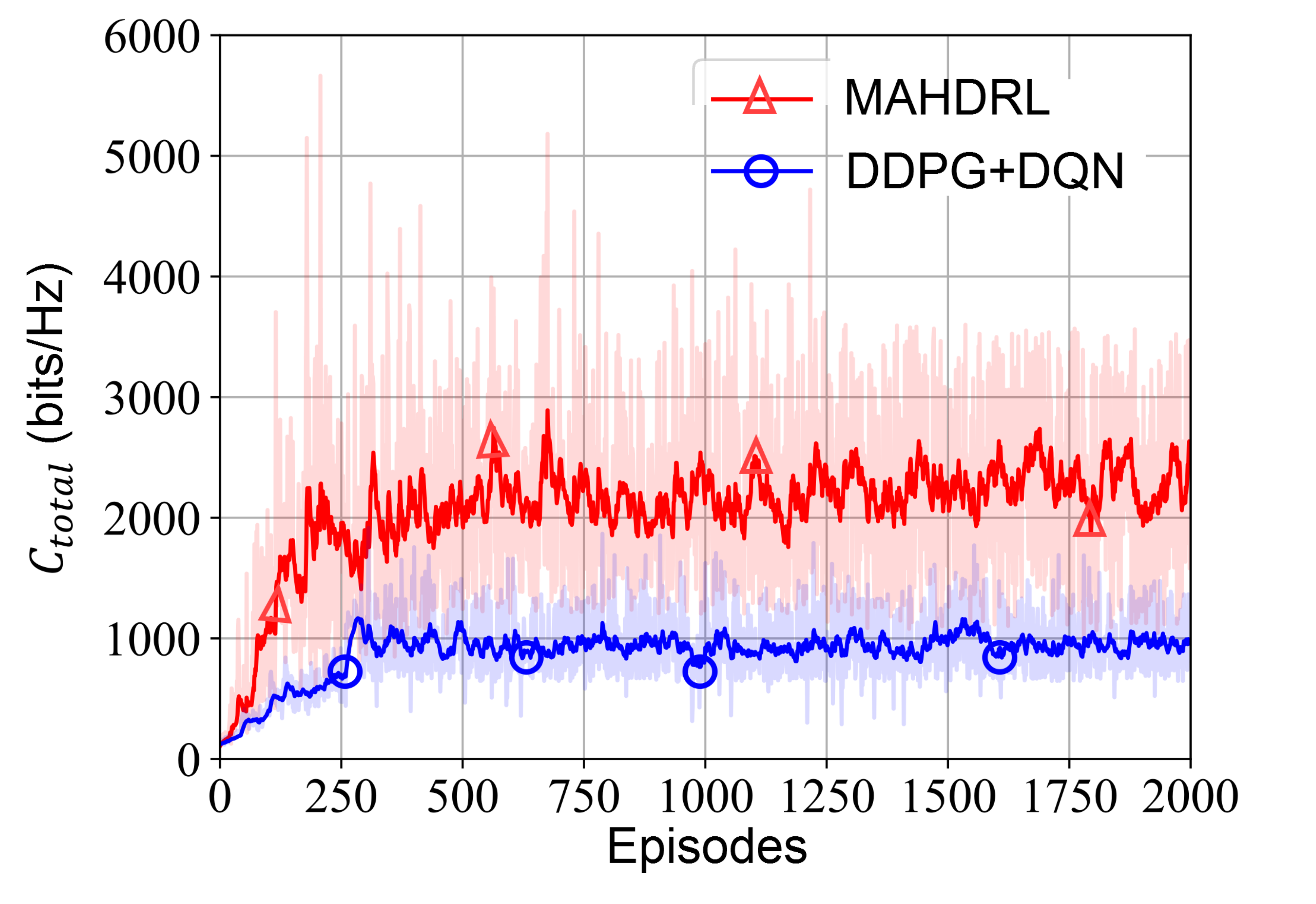}
		\vspace{-0.1in}
		\caption{Comparison of MAHDRL with DDPG+DQN.}
		\label{fig: MAHDRL performance in Traning}
		\vspace{-0.1in}
	\end{figure}
	
	\subsection{Comparison with Benchmarks}
	
	\subsubsection{Training stage}
	In the training stage, we compare the proposed MAHDRL with a benchmark, where the widely-used DDPG model in \cite{oubbati2022synchronizing} is adopted to replace the SAC in tier-1 in our proposed MAHDRL. For both approaches, the same DQN model as specified in Section \ref{section: DQN} is used in tier-2 for each UAV's WDC decisions. The SAC in the proposed MAHDRL and the DDPG in the benchmark have the same number of neural network layers and the same learning rate. Fig.~\ref{fig: MAHDRL performance in Traning} shows the WNs' total transmission data size $C_{total}$ over $2000$ trainings under both approaches. It is observed that our proposed MAHDRL with the SAC outperforms the benchmark with DDPG, due to the entropy-based loss function in the SAC that encourages the UAVs to explore the environment more. Although $C_{total}$ under the proposed MAHDRL approach fluctuates relatively wider than the benchmark, its fluctuation is still within an acceptable interval to achieve convergence.

	\subsubsection{Test stage}
	In the test stage, where the UAVs apply the well-trained policies to determine their actions, we compare our proposed MAHDRL solution with the following four different benchmark schemes:
	
	\begin{figure}[htbp]
		\setlength{\abovecaptionskip}{-0.0in}
		\centering
		\DeclareGraphicsExtensions{.eps,.mps,.pdf,.jpg,.png}
		\DeclareGraphicsRule{*}{eps}{*}{}
		\includegraphics[angle=0, width=0.45 \textwidth]{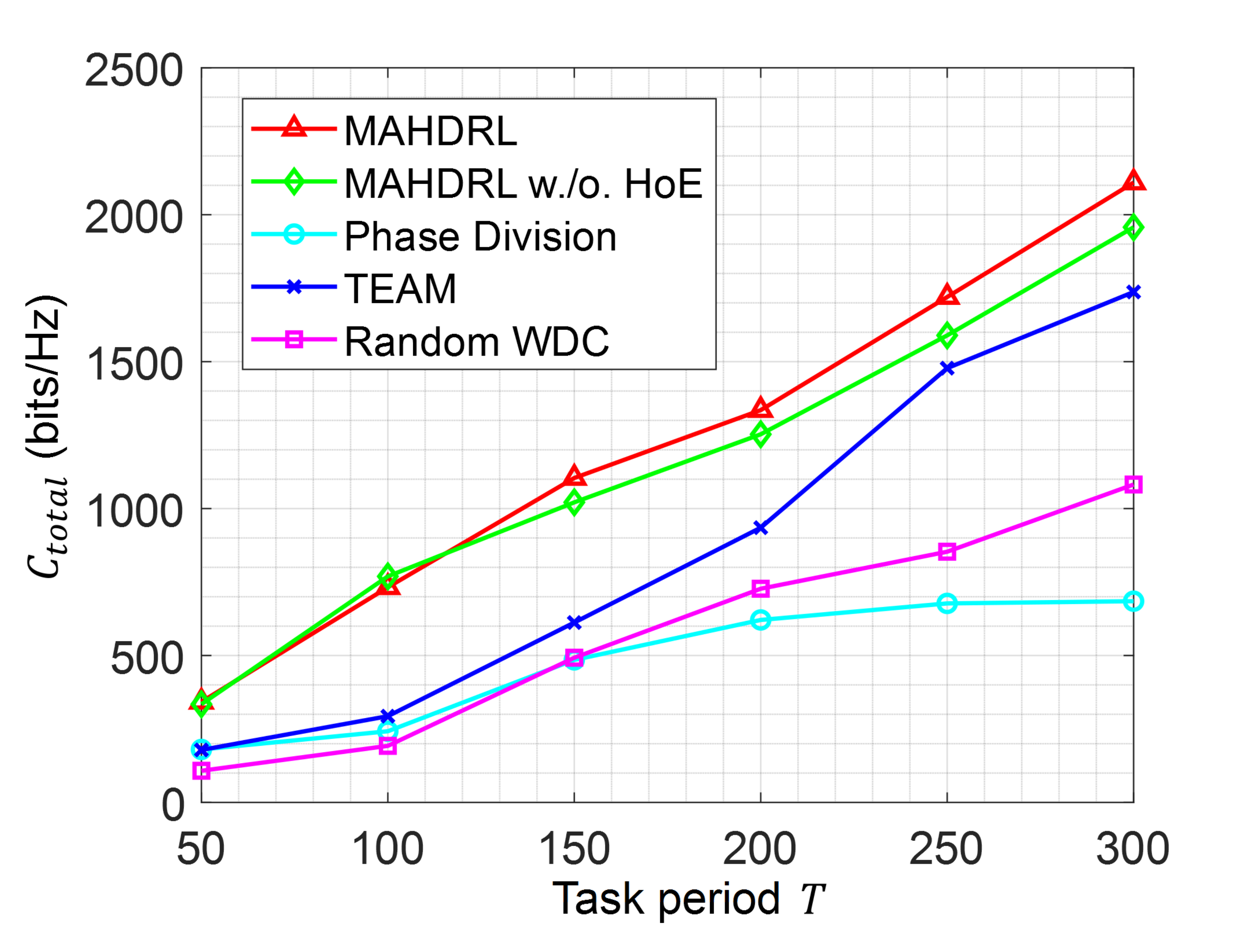}
		\vspace{-0.05in}
		\caption{Comparison of MAHDRL with benchmarks.}
		\label{fig: MAHDRL performance in Test}
	\end{figure}
	
	\textbf{(a) MAHDRL w./o. HoE}: Each E-node's HoE is not considered in this scheme, and the reward in (\ref{equ: reward for charge}) is reduced to $b_u^{WET}[t] \!=\! N_u[t] \!\cdot\! \sum_{w\in \mathcal{E}[t]}(B_w^u[t \!+\! 1] \!- \! B_w^u[t])$. All the other parts are the same as that in the proposed MAHDRL solution.
	\textbf{(b) Phase Division}: This benchmark scheme applies the classic time phase division for UAVs' WET and WDC as in \cite{xie2018throughput}. All the UAVs only transmit energy in the WET phase from slot $t \!=\!1$ to slot $t \!=\! \bar{T} \!-\! 1$, and only collect data from the I-nodes from slot $t=\bar{T}$ to slot $t=T$. We find use the optimal $\bar{T}$ that maximizes $C_{total}$ via one-dimensional exhaustive search.
	\textbf{(c) TEAM}: The TEAM scheme in \cite{oubbati2022synchronizing} is applied, where the UAVs are divided into two groups, one group is only responsible for WET and the other is only for WDC. We equally divide the UAVs into two groups.
	\textbf{(d) Random WDC}: The DQN in tier-2 is replaced by the random scheduling of the I-nodes over the sub-slots. For all the benchmark schemes, each UAV still applies the SAC algorithm to find its own trajectory, and for the former three benchmark schemes, each UAV applies the DQN algorithm to schedule the I-nodes for WDC.
	
	\begin{figure}[htbp]
		\setlength{\abovecaptionskip}{-0.0in}
		\centering
		\DeclareGraphicsExtensions{.eps,.mps,.pdf,.jpg,.png}
		\DeclareGraphicsRule{*}{eps}{*}{}
		\includegraphics[angle=0, width=0.45 \textwidth]{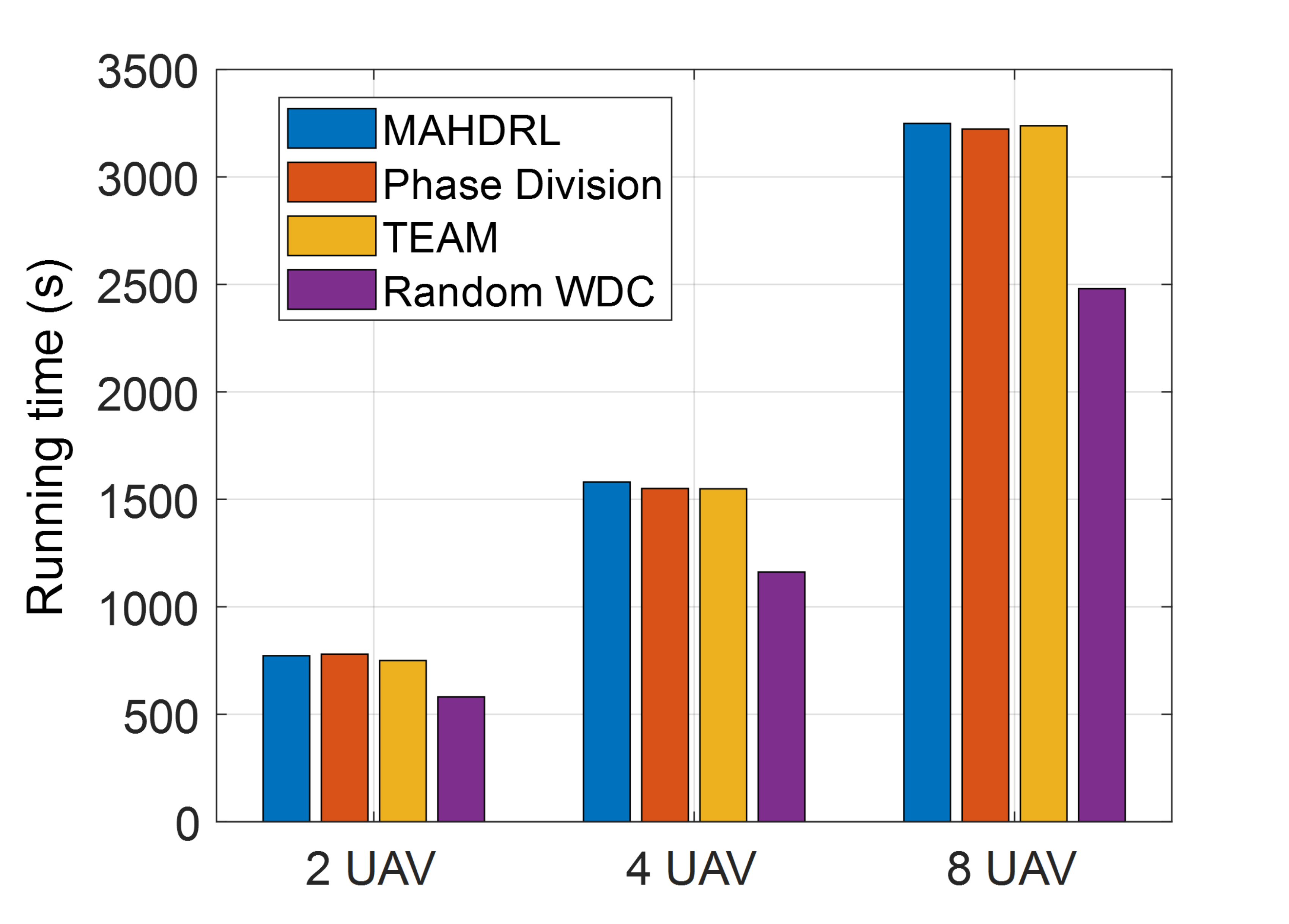}
		\caption{Running time comparison.}
		\label{fig: complexity}
	\end{figure}
	
	\begin{figure*}[h]
		\setlength{\abovecaptionskip}{-0.0in}
		\centering
		\setcounter{figure}{7}
		\DeclareGraphicsExtensions{.eps,.mps,.pdf,.jpg,.png}
		\DeclareGraphicsRule{*}{eps}{*}{}
		\includegraphics[angle=0, width=0.99 \textwidth]{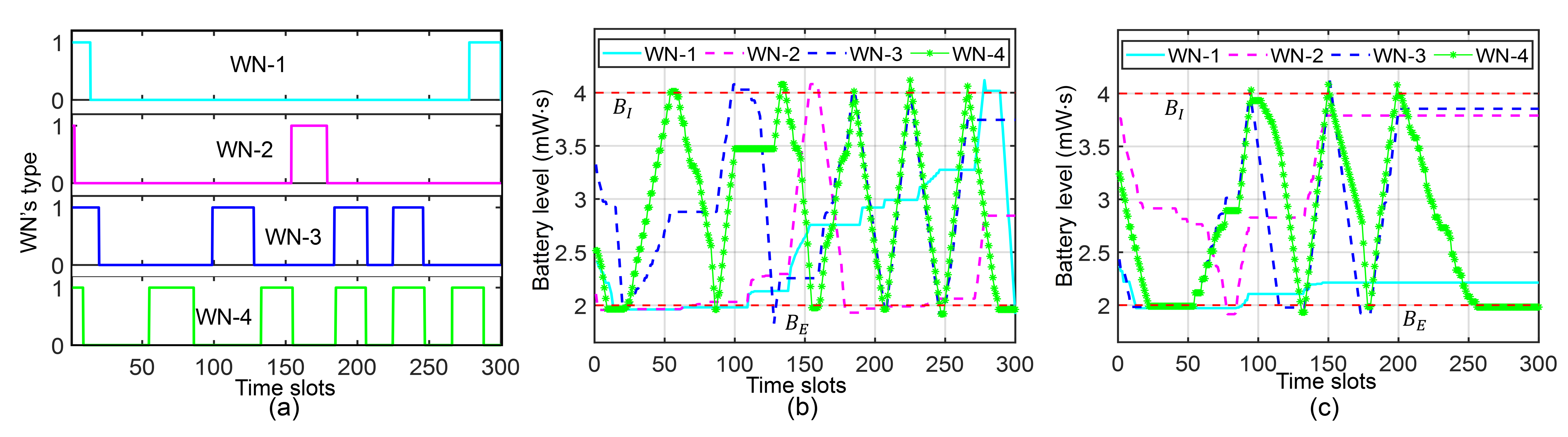}
		\vspace{-0.1in}
		\caption{Network example: (a) WNs' type variations. (b) WNs' battery levels under MAHDRL. (c) WNs' battery levels under MAHDRL w./o. HoE benchmark.}
		\vspace{-0.1in}
		\label{fig: MAHDRL performance in test}
	\end{figure*}
	
	First, Fig.~\ref{fig: MAHDRL performance in Test} shows the variations of $C_{total}$ over $T$ under the proposed and the above four benchmark schemes, respectively. It is observed that our proposed MAHDRL significantly outperforms all the benchmarks. This validates that by catering to the dynamic WN type updating over time, the performance of the multi-UAV aided WPCN can be largely improved. 
	
	Next, we compare the running time of the proposed MAHDRL approach with that of the benchmarks. It is easy to find from Section \ref{section: MAHDRL} that the MAHDRL w./o. HoE, the Phase Division and the TEAM have the same time complexity as our proposed MAHDRL approach, and the time complexity of the Random WDC approach is \small $\mathcal{O}\left(U \! \times \! EPS \! \times \! T \! \times \! \left(\sum_{j = 2}^{L_A}N_{A,j \!-\! 1} \! \cdot \! N_{A,j}\right.\right.$ $\left. \left. +\sum_{j = 2}^{L_V}N_{V,j \!-\! 1}\cdot N_{V,j}+\sum_{j = 2}^{L_Q}N_{Q, j \!-\! 1} \!\cdot\! N_{Q,j} \right) \right)$\normalsize without applying the DQN algorithm, which is less than our proposed MAHDRL approach and the other three benchmark approaches. In Fig.~\ref{fig: complexity}, since the MAHDRL w./o. HoE approach consumes almost the same running time as the proposed MAHDRL approach, we show the running time of the proposed MAHDRL, the Phase Division, the TEAM, and the Random WDC approaches under different UAV numbers. The running time of each approach is obtained as the average of $10$ simulations, where each simulation result is trained with 100 episodes. It is observed from Fig.~\ref{fig: complexity} that the running time of each approach almost doubles as the number of UAVs is doubled. Moreover, under each UAV number, the running time of the proposed MAHDRL is very close to that of the Phase Division and TEAM approaches, and the running time of the Random WDC approach is always the lowest. This is in accordance with our time complexity analysis, and further verifies the good performance of our proposed MAHDRL approach shown in Fig. \ref{fig: MAHDRL performance in Test}, as compared to the benchmarks.


	\subsection{Network Example}
	
	To further illustrate the performance of the proposed MAHDRL scheme, we consider a network of a smaller scale, with $2$ UAVs and $4$ WNs randomly locating within a horizontal area of $300$ m$\times300$ m. Fig.~\ref{fig: UAV trajectories} shows the UAVs' trajectories and WET decisions over time. The shortest distance between the two UAVs in Fig.~\ref{fig: UAV trajectories} is $2.74$ m, which is larger than the required safe distance $d_{min}$. It is also observed that each UAV does not always select $Z_u[t]=1$ to save its energy.
	
	\begin{figure}[htbp]
		\setlength{\abovecaptionskip}{-0.0in}
		\centering
		\setcounter{figure}{6}
		\DeclareGraphicsExtensions{.eps,.mps,.pdf,.jpg,.png}
		\DeclareGraphicsRule{*}{eps}{*}{}
		\includegraphics[angle=0, width=0.45 \textwidth]{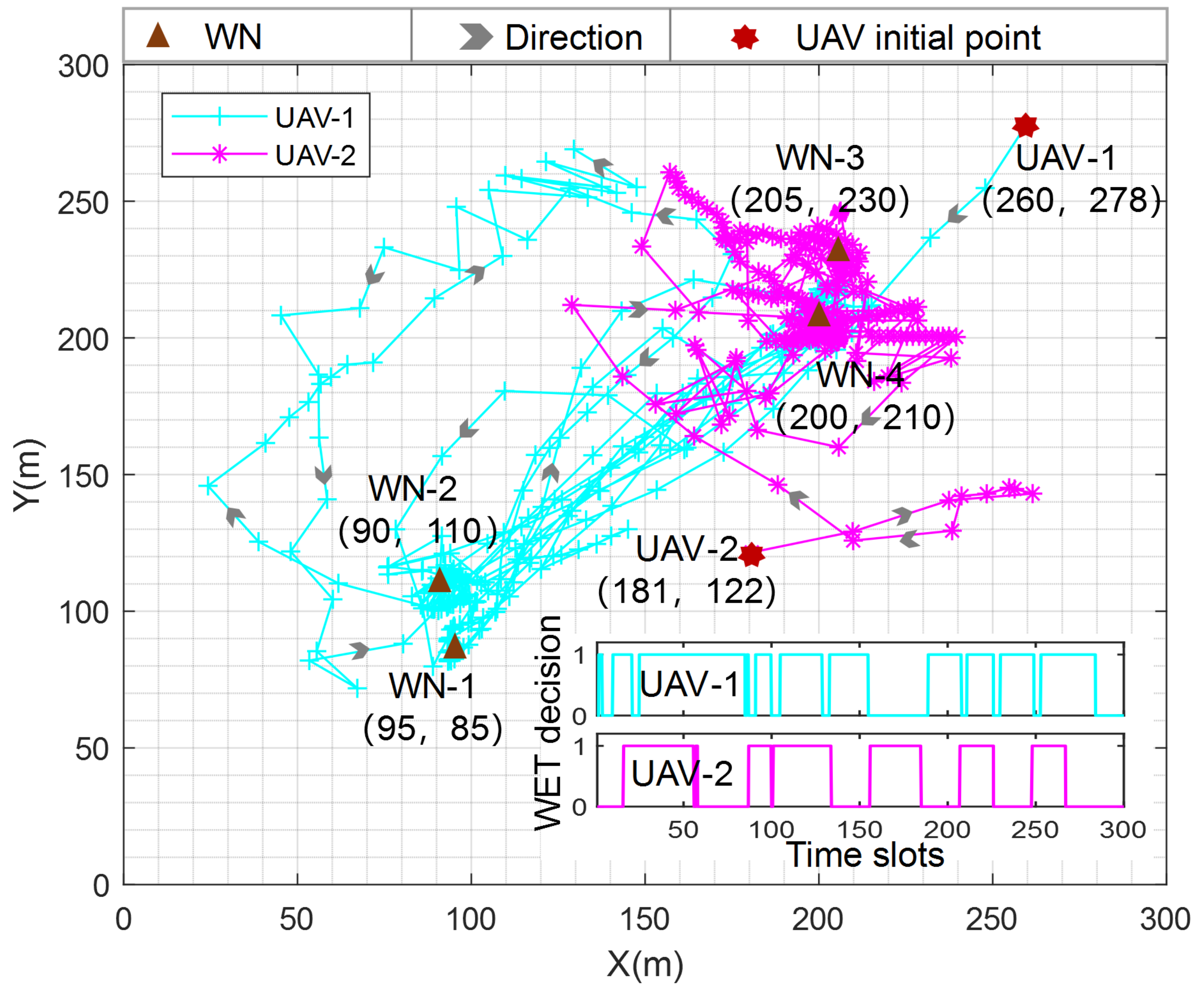}
		\vspace{-0.05in}
		\caption{UAVs' trajectories and WET decisions.}
		\label{fig: UAV trajectories}
	\end{figure} 
	
	Fig.~\ref{fig: MAHDRL performance in test}~(a) and Fig.~\ref{fig: MAHDRL performance in test}~(b) show each WN's type and battery energy variations over time, respectively, where each WN's battery energy increases when it is an E-nodes, or decreases when it is an I-node. The WN type updating under the two thresholds $B_I$ and $B_E$ follows (\ref{equ: wireless node flag update}). It is also observed from Fig.~\ref{fig: UAV trajectories} that since only UAV-$1$ comes to WN-$1$ and WN-$2$ for WET, they have fewer chances to become I-nodes than WN-$3$ and WN-$4$, where both UAVs fly to them and transmit energy. However, once WN-$1$ and WN-$2$ become I-nodes, they have higher chances to be scheduled to transmit data. The total transmission data size (bits/Hz) of the $4$ WNs in the $T$ slots is $263.42$, $487.01$, $638.49$ and $733.88$, respectively, satisfying the constraint in (\ref{constraint: a}). It is also interesting to observe similar battery charging rates at WN-3 and WN-4 from time slot $176$ to slot $268$ in Fig.~\ref{fig: MAHDRL performance in test}~(b). This is mainly because that, as observed from Fig.~\ref{fig: UAV trajectories}, UAV-$1$ or UAV-$2$ flies in the neighborhood of WN-$3$ and WN-$4$ from slot $176$ to slot $268$ or slot $17$ to slot $268$, respectively. During the corresponding time slots, each UAV transmits energy to WN-$3$ and WN-$4$ from sufficiently short distances, such that (close to) saturated power is harvested at WN-$3$ and WN-$4$, which leads to their similar battery energy charging rates. Also, since the closely located UAV-$2$ can efficiently collect data from both WN-$3$ and WN-$4$, the battery energy decreasing rates are also similar.
	
	We also show each WN's battery energy under the MAHDRL w./o. HoE benchmark in Fig.~\ref{fig: MAHDRL performance in test}~(c). It is observed that unlike Fig.~\ref{fig: MAHDRL performance in test}~(b), where each WN can become an I-node with its battery energy exceeding $B_I$, WN-$1$ and WN-$2$ in Fig.~\ref{fig: MAHDRL performance in test}~(c) cannot harvest sufficient energy over all $T$ slots and thus cannot meet the constraint in (\ref{constraint: a}). This matches with Fig.~\ref{fig: MAHDRL performance in Test}, where our proposed MAHDRL outperforms the MAHDRL w./o. HoE benchmark. In addition, at the end of $T=300$ slots, the remaining battery energy of the two UAVs are $72399.97$ W$\cdot$s and $30115.61$ W$\cdot$s, respectively, satisfying the constraint in (\ref{constraint: c}).
	
	\begin{figure}[htbp]
		\setlength{\abovecaptionskip}{-0.0in}
		\centering
		\setcounter{figure}{8}
		\DeclareGraphicsExtensions{.eps,.mps,.pdf,.jpg,.png}
		\DeclareGraphicsRule{*}{eps}{*}{}
		\includegraphics[angle=0, width=0.45 \textwidth]{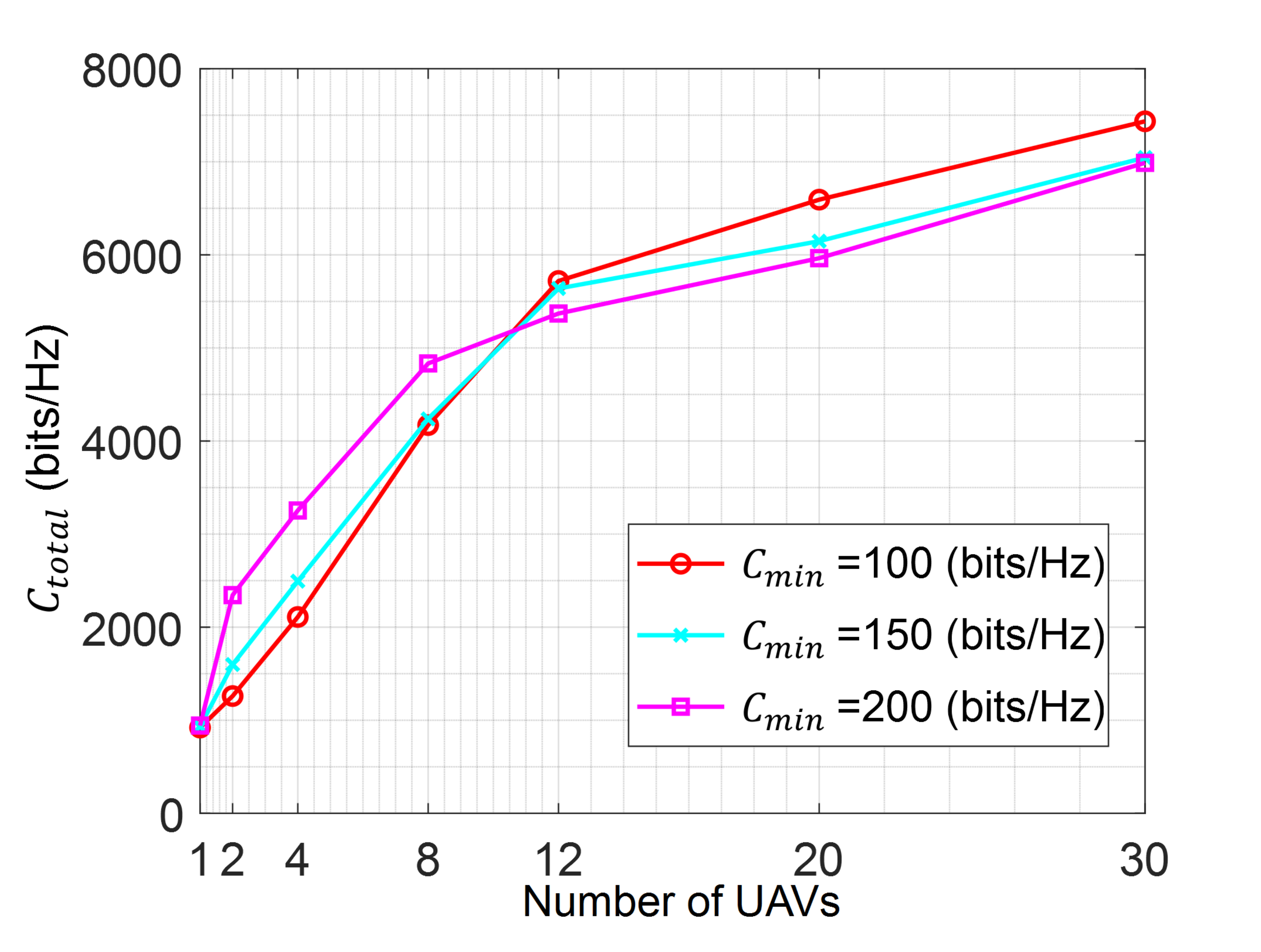}
		\vspace{-0.05in}
		\caption{Impact of UAV number on $C_{total}$.}
		\label{fig: scalability}
	\end{figure}
	
    Next, in Fig.~\ref{fig: scalability}, to show the scalability of our proposed MAHDRL approach, we gradually increase the network scale and show the achieved $C_{total}$ under different $C_{min}$, where we consider that the number of WNs always equals the twice of the number of UAVs, i.e. $W = 2U$. When the number of the UAVs increases from $U=1$ to $U=30$ in Fig.~\ref{fig: scalability}, the number of WNs increases from $W=2$ to $W=60$ accordingly. It is observed from Fig.~\ref{fig: scalability} that when  $U=1$, the same $C_{total}$ is achieved under different $C_{min}$, and when  $1<U<12$, a larger $C_{total}$ is always achieved under a larger $C_{min}$ under each value of $U$, since most of the E-nodes can be properly transformed into I-nodes to meet $C_{min}$; however, when $U\geq 12$, due to the increased information transmission interference under the large UAV number to meet a large $C_{min}$, it is found that a larger $C_{total}$ is achieved under a smaller $C_{min}$ over each value of $U$. It is also observed that under each $C_{min}$, the value of $C_{total}$ always increases over the UAV number and thus the network scale. This verifies the scalability of our proposed MAHDRL approach. 
	
	\begin{figure}[htbp]
		\setlength{\abovecaptionskip}{-0.0in}
		\centering
		\DeclareGraphicsExtensions{.eps,.mps,.pdf,.jpg,.png}
		\DeclareGraphicsRule{*}{eps}{*}{}
		\includegraphics[angle=0, width=0.45 \textwidth]{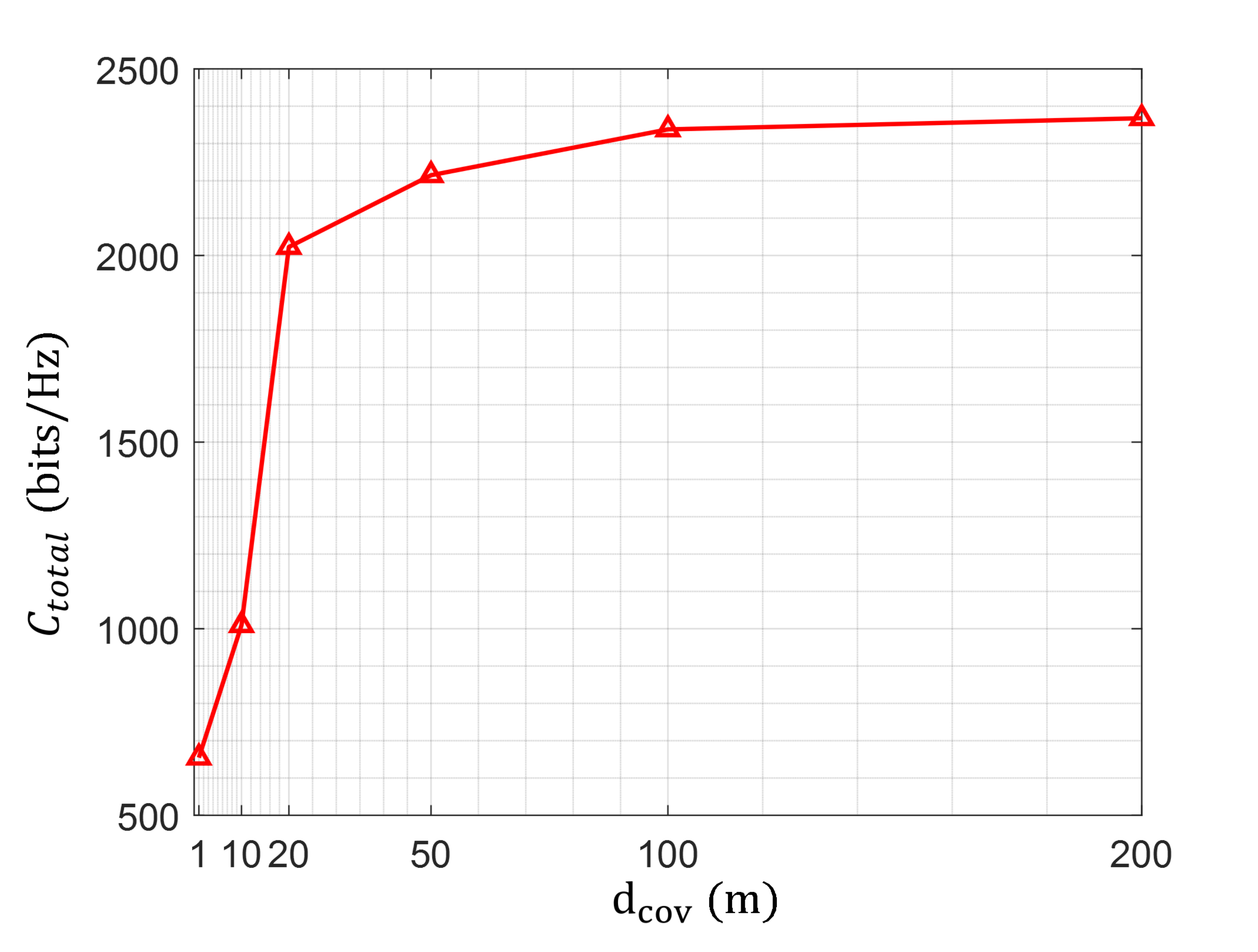}
		\vspace{-0.06in}
		\caption{Impact of $d_{\rm{cov}}$ on $C_{total}$.}
		\label{fig: d_cov}
		\vspace{-0.06in}
	\end{figure}
	
	At last, Fig.~\ref{fig: d_cov} shows the impact of the status reporting distance $d_{\rm{cov}}$ in (\ref{equ: uav knows battery}) on $C_{total}$. It is observed from Fig.~\ref{fig: d_cov} that as $d_{cov}$ increases, more information of the  E-nodes can be obtained at the central trainer, and thus $C_{total}$ increases; and when $d_{cov}$ is sufficiently large, such that almost each E-node's complete status can be all obtained by the central trainer, the POMDP modeled in Sections \ref{subsection: sac pomdp} and \ref{section: DQN pomdp} approaches an MDP, respectively, where $C_{total}$ is almost unchanged.

	\section{Conclusion} \label{section: conclusion}
	
	This paper proposed a novel design of the multi-UAV aided WPCN with repeatedly-changing WN types over time. By applying the LoS-probability based A2G channels, we utilized the practical non-linear energy harvesting model at each E-node, and further developed all the UAVs' and the WNs' battery energy management models. To effectively solve the complicated total transmission data size maximization problem, the new MAHDRL framework with two tiers was proposed. We designed the central training of the SAC policy and the local training of the DQN policy by exploiting the interactions between the UAVs and the WNs. Once well trained, both the SAC policy and the DQN policy are executed distributivity at each UAV. Extensive simulations validated that our proposed MAHDRL approach that adapts to the network dynamics outperforms benchmarks. In practice, the deployment of the central trainer and its available computation resources may largely affect the performance of the multi-UAV aided WPCN. In the future work, it is interesting to study the UAV-enabled mobile central trainer and investigate the joint transmission and computation design of the multi-UAV aided WPCN.

	\bibliographystyle{IEEEtran}
	\bibliography{references}{}
	

\end{document}